\documentclass[
    a4paper,
    amsfonts,
    amssymb,
    amsmath,
    reprint,
    floatfix,
    showkeys,
    nofootinbib,
    pra,
    aps
]{revtex4-2}

\usepackage[english]{babel}
\usepackage[utf8]{inputenc}
\usepackage{amsthm}
\usepackage{mathtools}
\usepackage{multirow}
\usepackage{physics}
\usepackage{xcolor}
\usepackage{booktabs}
\usepackage{graphicx}
\usepackage{adjustbox}
\usepackage{listings}
\usepackage[T1]{fontenc}
\usepackage{csquotes}
\usepackage{xurl}
\usepackage[normalem]{ulem}
\usepackage[
    colorlinks=true,
    linkcolor=blue,
    citecolor=blue,
    urlcolor=blue
]{hyperref}

\begin{document}
\title{Quantum Circuits for Quantum Spatial Search on \texorpdfstring{$d$}{d}-Dimensional Hypercubic Lattices}

\author{Rei Sato}
    \email[Correspondence email address: ]{rei@classiq.io}
    \affiliation{Classiq Technologies G.K.,
  1-5-1 Marunouchi, Chiyoda-ku, Tokyo, 100-6509, Japan}
\date{\today}

\begin{abstract}
We propose an explicit quantum-circuit construction for quantum spatial search based on discrete-time quantum walks on $d$-dimensional hypercubic lattices and systematically evaluate its circuit-resource scaling.  Although the algorithmic performance of quantum spatial search has been extensively studied, a systematic evaluation of its circuit-resource scaling remains limited.  We implement the flip-flop shift operator using coin-controlled modular increment and decrement operations for position updates and a single $X$ gate for direction reversal.  We compare four implementations of the position update based on the quantum Fourier transform (QFT), ripple-carry addition, multi-controlled $X$ (MCX) gates, and automatically synthesized arithmetic.  The results show that the QFT-based implementation reduces circuit width and depth, while the ripple-carry implementation reduces the CX-gate count at the cost of additional ancilla qubits.  We also extend the construction to non-periodic boundary conditions by introducing self-loops at lattice boundaries.  We confirm that numerical simulations of the circuits reproduce the theoretical search dynamics for $d=2,3,4$, and $5$ under periodic boundary conditions and for $d=2$ under non-periodic boundary conditions.  For $N=L^d$ lattice sites, the circuit width scales as $\mathcal{O}(\log N)$.  We derive asymptotic upper bounds on the total circuit depth of $\mathcal{O}(\sqrt{N}\log^{5/2} N)$ for $d=2$ and $\mathcal{O}(\sqrt{N}\log^2 N)$ for $d\geq3$, consistent with the numerical results.  Furthermore, CX-gate depolarizing noise simulations show that reducing the CX-gate count improves noise robustness.  These results provide guidance for efficient quantum-circuit design for spatial search on hypercubic lattices.
\end{abstract}


\maketitle

\section{Introduction}

Quantum walks use quantum superposition and interference to achieve faster propagation than classical random walks under suitable conditions~\cite{10.5555/1070432.1070590,PhysRevA.48.1687}. Quantum walks have also been experimentally realized on several physical platforms, including trapped-ion and photonic systems~\cite{PhysRevLett.103.090504,peruzzo2010quantum}.

One application of quantum walks is quantum spatial search, which aims to find a marked site on a graph by amplifying its probability amplitude. The optimal number of search steps, at which the success probability reaches its first peak, depends strongly on the structure and dimension of the underlying graph. For a $d$-dimensional hypercubic lattice with $N$ sites, in which each site is connected to its $2d$ nearest neighbors along the coordinate axes, it scales as $\mathcal{O}(N)$ for $d=1$, where no quantum speedup over classical search is obtained~\cite{10.5555/1070432.1070590}, as $\mathcal{O}(\sqrt{N\log N})$ for $d=2$~\cite{10.5555/1070432.1070590,PhysRevA.78.012310}, and as $\mathcal{O}(\sqrt{N})$ for $d\geq3$~\cite{10.5555/1070432.1070590,J2018,PhysRevA.82.032330}.  Quantum spatial search has also been investigated on fractal lattices, where numerical studies indicate that the optimal number of search steps scales as $\mathcal{O}(N^{1/d_s})$ for $d_s<2$ and as $\mathcal{O}(\sqrt{N})$ for $d_s>2$, where $d_s$ is the spectral dimension~\cite{PhysRevA.86.012332,PhysRevA.101.022312}.

More recently, possible applications of quantum spatial search to image processing and combinatorial optimization have also been explored~\cite{giri2025quantum,PhysRevResearch.2.023302}. These developments motivate explicit circuit-level realizations of quantum spatial search.

A single step of quantum spatial search~\cite{10.5555/1070432.1070590} typically consists of a coin operation, a flip-flop shift, and an oracle that marks the target site. The flip-flop shift moves the walker to a neighboring site according to the direction label encoded in the coin register and reverses the corresponding direction label after the move. Consequently, the shift operator directly encodes the adjacency structure of the graph, and the position update and direction-label reversal must be implemented consistently.

For the Boolean hypercube graph $\{0,1\}^n$, which should be distinguished from the hypercubic lattices $\{0,\ldots,L-1\}^d$ considered here, this difficulty is less apparent. The shift operation moves the walker to the neighboring site associated with the selected direction, and this transformation itself realizes the flip-flop shift without requiring a separate direction-label reversal operation~\cite{PhysRevA.67.052307, WingBocanegra2025}.  By contrast, for general lattice walks, the direction-label reversal is a nontrivial operation that must be implemented together with the position update.

A related study formulated two-dimensional quantum-walk search with both open and periodic boundary conditions in the context of ordered marked nodes and tracking problems~\cite{Sahu2024}.  The position update in such circuits can be implemented with different arithmetic constructions, including QFT-based adders~\cite{shakeel2020efficient,draper2000addition,koch2022gate}, MCX gates~\cite{puengtambol2021implementation,Sahu2024,douglas2009efficient}, and ripple-carry adders~\cite{cuccaro2004new}. Since the choice of construction affects the circuit width, gate count, and depth, a unified comparison of these constructions within the overall search circuit is important for practical circuit design.  It also proposed circuit constructions based on coin-controlled increment and decrement operations and coin-state reversal. However, an explicit reversible circuit that coherently evaluates boundary-dependent move validity and conditionally implements the resulting non-periodic shift was not provided.  This motivates the development of a general circuit-level framework for periodic $d$-dimensional hypercubic lattices, together with numerical validation and systematic resource and noise analyses of the synthesized circuits.

In this work, we develop an explicit circuit construction of quantum spatial search on $d$-dimensional hypercubic lattices.  For periodic lattices, we encode the $2d$ direction labels so that each pair of opposite directions along a spatial axis differs only in the least significant bit of the coin register.  This encoding allows the direction-label reversal to be implemented by a single $X$ gate, while the position update is realized by coin-controlled modular increment and decrement operations on the corresponding position register.

We also present an explicit circuit construction for two-dimensional lattices with non-periodic boundary conditions. In this case, the shift operation is conditioned on the current position so that transitions beyond the lattice boundary are blocked.

We implement the proposed circuit using Qmod~\cite{vax2025qmod}.  We verify that the generated circuits reproduce the theoretical search dynamics on periodic hypercubic lattices with $d=2,3,4$, and $5$ and on a two-dimensional lattice with non-periodic boundary conditions. We then evaluate the scaling of the circuit width and depth and examine the influence of the initial-state preparation method. Finally, we compare default and CX-optimized circuits under a CX-gate depolarizing noise model.

The main contributions of this work are as follows.  First, we construct the flip-flop shift operator for periodic $d$-dimensional hypercubic lattices, in which the direction-label reversal is implemented by a single $X$ gate. Second, we present and validate a position-dependent shift construction for a $d=2$ hypercubic lattice with non-periodic boundary conditions. Third, we compare four implementations of the position update in terms of circuit width, CX-gate count, and depth. Fourth, we derive asymptotic upper bounds on the circuit depth, consistent with the numerical results for $d=2$--$5$, and evaluate the robustness against CX-gate depolarizing noise.

The remainder of this paper is organized as follows. Section~\ref{sec:quantum_spatial_search} introduces the mathematical formulation of quantum spatial search based on discrete-time quantum walks. Section~\ref{sec:quantum_circuit} presents the proposed circuit construction for $d$-dimensional hypercubic lattices. Section~\ref{sec:experiment} reports the numerical validation, circuit-resource analysis, and noise simulations. Section~\ref{sec:discussion} discusses the implications and limitations of the results, and Sec.~\ref{sec:conclusion} concludes the paper.

\section{Quantum Spatial Search}
\label{sec:quantum_spatial_search}

\begin{figure}[ht] 
    \centering
    \includegraphics[width=1.0\linewidth]{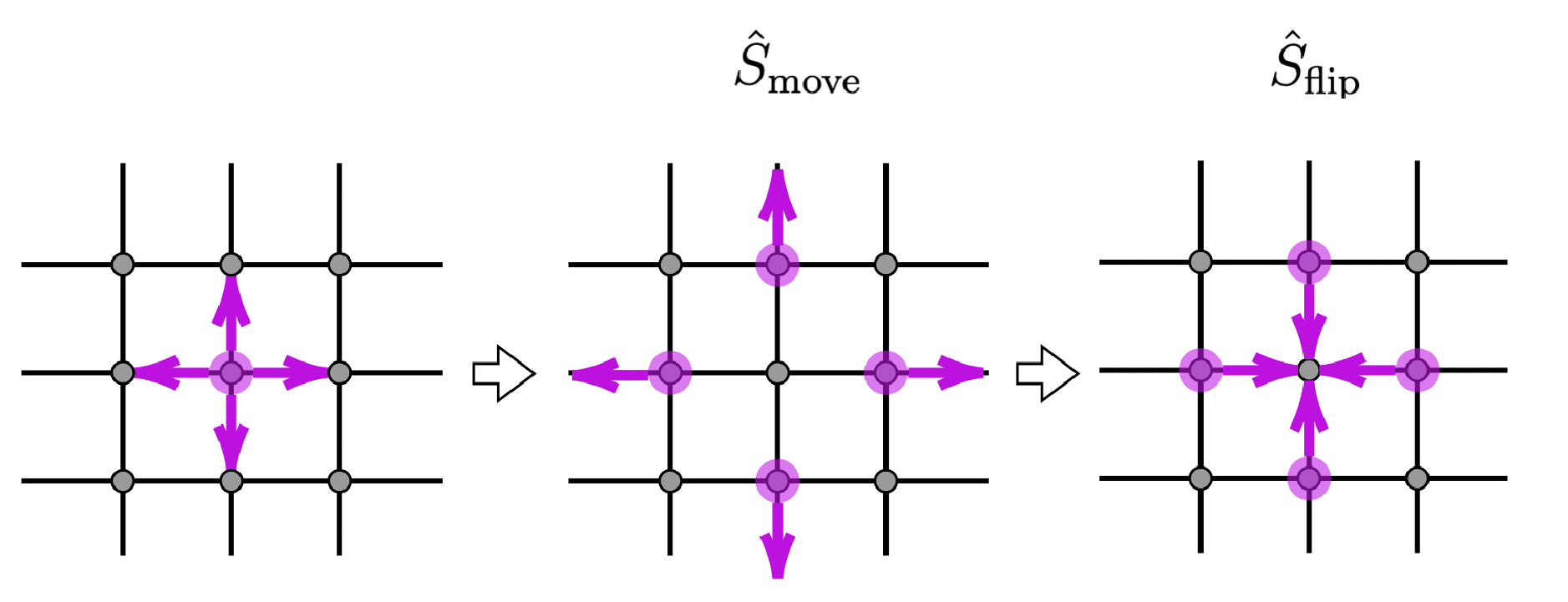}
    \caption{Definition of the flip-flop walk on a two-dimensional hypercubic lattice. }
    \label{fig:flip-flop-walk}
\end{figure}

Quantum spatial search based on a discrete-time quantum walk is formulated on a Hilbert space $\mathcal{H} = \mathcal{H}_x \otimes \mathcal{H}_C$. We consider a $d$-dimensional hypercubic lattice of linear size $L$ with periodic boundary conditions, that is, a graph with site set $V = \{0,\ldots,L-1\}^d$ in which each site is connected to its $2d$ nearest neighbors along the coordinate axes, with each coordinate taken modulo $L$. Each site is represented by a position vector $\mathbf{x} = (x_0,\ldots,x_{d-1}) \in V$, and the total number of sites is $N = L^d$. Let $\ket{\mathbf{x}}$ denote the basis states of the position space $\mathcal{H}_x$, and let $\ket{l}$ represent the basis states of the coin space $\mathcal{H}_C$.  

For a $d$-dimensional hypercubic lattice, the degree of the site is $k=2d$, and the coin state $l=0,\ldots,k-1$ represents the direction of motion to a neighboring site. 
The state of the quantum spatial search at step $t$, $\ket{\Psi(t)}$, is described as
\begin{equation}
    \ket{\Psi(t)} = \sum_{\mathbf{x} \in V} \sum_{l=0}^{k-1} \psi_{\mathbf{x}, l}(t) \ket{\mathbf{x}} \otimes \ket{l},
    \label{eq:init_state}
\end{equation}
where $\psi_{\mathbf{x}, l}(t)$ are the probability amplitudes associated with position $\mathbf{x}$ and coin state $l$.

The discrete-time evolution of the quantum walker is governed by the operators,
\begin{equation}
    \ket{\Psi(t)} = [\hat{W}\hat{R}]^t \ket{\Psi(0)}.
    \label{eq:time_evolv}
\end{equation}
Here, $\hat{R}$ is the oracle operator that marks the target site $\mathbf{x}_0$ by applying a phase flip, and is defined as
\begin{equation}
    \hat{R}
    =
    \left(
        \hat{I}_x
        -
        2\ket{\mathbf{x}_0}\bra{\mathbf{x}_0}
    \right)
    \otimes\hat{I}_C.
\end{equation}

The quantum walk operator $\hat{W}$ comprises the coin operator $\hat{C}$ and the shift operator $\hat{S}$,
\begin{equation}
    \hat{W}=\hat{S}\hat{C}.
\end{equation}

The coin operator,
\begin{equation}
    \hat{C} = \hat{I}_{\mathbf{x}} \otimes \hat{G},
\end{equation}
acts on the internal degrees of freedom of the walker at each site.  In this work, the coin operator $\hat{G}$ is chosen as the Grover diffusion operator,
\begin{equation}
    \hat G = 2|s\rangle\langle s| - \hat I_C,
    \label{eq:grover_diffuser}
\end{equation}
where
\begin{equation}
    \ket{s} = \frac{1}{\sqrt{k}}\sum_{l=0}^{k-1} \ket{l}
\end{equation}
is the uniform superposition over the coin states.

The shift operator $\hat{S}$, known as the flip-flop shift shown in Fig.~\ref{fig:flip-flop-walk}, moves the walker to an adjacent site and then reverses its internal direction.  It is defined by
\begin{equation}
    \hat{S} \left(\ket{\mathbf{x}} \otimes \ket{l}\right) 
    = \ket{\mathbf{x} + \boldsymbol{\delta}_l} \otimes \ket{\bar{l}},
    \label{eq:flip-flop_walk}
\end{equation}
where $\boldsymbol{\delta}_l$ denotes the unit displacement vector along one of the coordinate axes associated with the coin state $l$, with the lattice spacing set to unity, and $\bar{l}$ denotes the reversed coin state.

For instance, for a two-dimensional square lattice ($d=2$), the coin states $l \in \{0,1,2,3\}$ are mapped to the cardinal directions $\{-x, +x, -y, +y\}$.  
The displacement vectors are then $\boldsymbol{\delta}_0=(-1,0)$, $\boldsymbol{\delta}_1=(+1,0)$, $\boldsymbol{\delta}_2=(0,-1)$, and $\boldsymbol{\delta}_3=(0,+1)$ for $(x,y)$.  In the flip-flop shift, each coin state is mapped to the opposite direction after the position update, e.g., $l=0$ to $\bar{l}=1$ and $l=2$ to $\bar{l}=3$.

Finally, the probability distribution of finding the walker at site $\mathbf{x}$ at time $t$ is given by
\begin{equation}
    P(\mathbf{x}, t) = \sum_{l=0}^{k-1} |\langle \mathbf{x}, l | \Psi(t) \rangle|^2.
    \label{eq:time_prob}
\end{equation}

\section{Quantum Circuit}
\label{sec:quantum_circuit}
We implement quantum spatial search circuits using Qmod~\cite{vax2025qmod}.  We primarily consider discrete-time quantum walks on periodic $d$-dimensional hypercubic lattices, as formulated in Sec.~\ref{sec:quantum_spatial_search}.  We also extend the circuit construction to non-periodic boundary conditions.

\subsection{Periodic boundary condition}
\begin{figure}[ht] 
    \centering
    \includegraphics[width=1.0\linewidth]{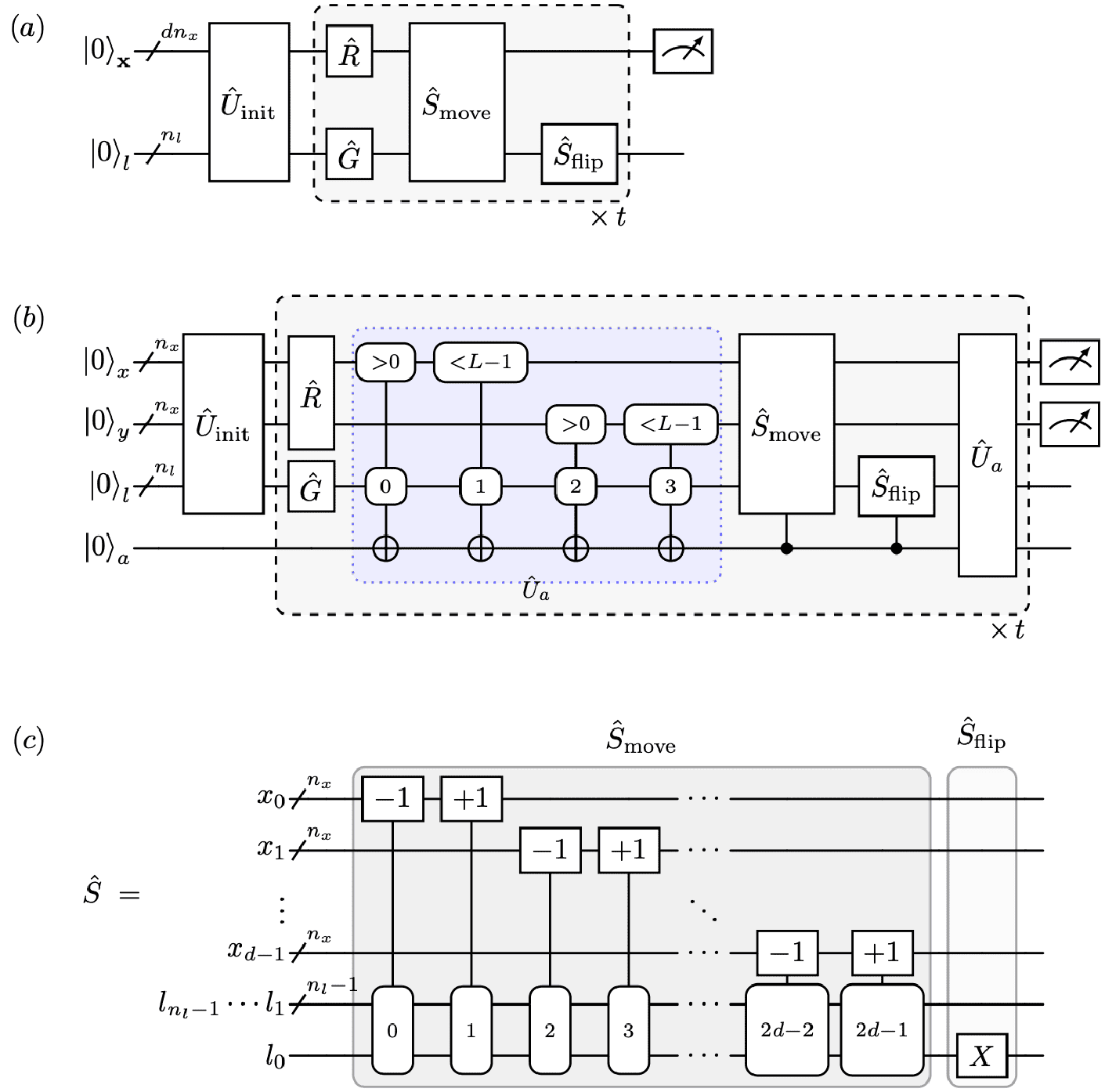}
    \caption{
    Quantum circuits for quantum spatial search by a discrete-time quantum walk. (a) Circuit on a $d$-dimensional hypercubic lattice with periodic boundary conditions. (b) Circuit on a $d=2$ lattice with non-periodic boundary conditions. The operator $\hat{U}_a$ in dotted box computes the validity flag $a(\mathbf{x},l)$ defined in Eq.~\eqref{eq:validity-flag-2d} into the auxiliary qubit, and the shift is applied only when the flag is $1$. A second application of $\hat{U}_a$ to the updated state returns the auxiliary qubit to $\ket{0}_a$. (c) Circuit implementation of $\hat{S}$. In $\hat{S}_{\rm move}$, the position register $x_j$ is decremented modulo $L$ when the coin state is $\ket{2j}$ and incremented when it is $\ket{2j+1}$, where the rounded nodes indicate the controlling coin values. $\hat{S}_{\rm flip}$ applies an $X$ gate to the least significant coin qubit $l_0$.
    }
    \label{fig:proposed_circuit}
\end{figure}

Figure~\ref{fig:proposed_circuit}(a) shows the proposed quantum circuit for spatial search. The initial state is prepared as a uniform superposition over the position and coin states:
\begin{align}
|\Psi(0)\rangle
    &= U_{\rm init}|0\rangle^{\otimes d n_x}|0\rangle^{\otimes n_l} \notag\\
    &= \frac{1}{\sqrt{Nk}}\sum_{\mathbf{x}\in V}\sum_{l=0}^{k-1}|\mathbf{x}\rangle\otimes|l\rangle,
\label{eq:initial_state_valid}
\end{align}
where $k=2d$, $n_x=\lceil\log_2 L\rceil$ is the number of qubits per spatial coordinate, and $n_l=\lceil\log_2 k\rceil$ is the number of qubits for the coin state.  When $L=2^{n_x}$ and $k=2^{n_l}$, the initial state can be prepared using Hadamard gates.  Otherwise, $U_{\rm init}$ prepares a uniform superposition over the valid position and coin states, with zero amplitude assigned to unused computational-basis states.

For $d=2$ and $4$, the coin registers can be initialized using Hadamard gates, since $k=2d$ is a power of two.  For $d=3$ and $5$, the coin registers contain two and six unused computational-basis states, respectively, whose amplitudes are set to zero by $U_{\rm init}$.  Similarly, when $L\neq 2^{n_x}$, unused position states are assigned zero amplitude.

For the oracle operator, we choose the marked site at the center of the lattice, $\mathbf{x}_0=(\lfloor L/2\rfloor,\ldots,\lfloor L/2\rfloor)$.  The oracle applies a phase flip to all coin states at the marked site, $\hat{R}(|\mathbf{x}_0\rangle\otimes|l\rangle)=-|\mathbf{x}_0\rangle\otimes|l\rangle$.

The coin operator is implemented using the Grover diffusion operator defined in Eq.~\eqref{eq:grover_diffuser}, $\hat{G}=\hat{U}_s(2|0\rangle\langle0|^{\otimes n_l}-\hat{I}_C)\hat{U}_s^\dagger$, where $\hat{U}_s|0\rangle^{\otimes n_l}=|s\rangle$ prepares the uniform superposition over the $k=2d$ coin states.

For the shift operator $\hat{S}$ defined in Eq.~\eqref{eq:flip-flop_walk}, we assign the coin state $|l\rangle$ with $l=2j$ and $l=2j+1$ to the negative and positive directions along each spatial axis $j\in\{0,\ldots,d-1\}$, respectively. Since these states differ only in the least significant bit, the flip-flop shift can be decomposed into a position update $\hat{S}_{\rm move}$ followed by a direction reversal $\hat{S}_{\rm flip}$:
\begin{equation}
    \hat{S} =(\hat{I}_x\otimes\hat{S}_{\rm flip})\hat{S}_{\rm move}.
\end{equation}

The move operator updates the position register according to the coin state,
\begin{equation}
    \hat{S}_{\rm move}
    (\ket{\mathbf{x}}\otimes\ket{l})
    =
    \ket{\mathbf{x}+\boldsymbol{\delta}_l}
    \otimes\ket{l},
    \label{eq:shift-op}
\end{equation}
where $\boldsymbol{\delta}_{2j}$ and $\boldsymbol{\delta}_{2j+1}$ denote the displacements by one lattice site along the negative and positive directions of axis $j$, respectively.  The position update is implemented using coin-controlled modular increment and decrement operations, as shown in Fig.~\ref{fig:proposed_circuit}(c).

The flip operator reverses the coin direction by applying a single $X$ gate to the least significant coin qubit:
\begin{equation}
    \hat{S}_{\rm flip}
    =
    \hat{I}^{\otimes(n_l-1)}
    \otimes\hat{X}.
    \label{eq:sflip_x_gate}
\end{equation}
Thus, $\hat{S}_{\rm flip}\ket{2j}=\ket{2j+1}$ and $\hat{S}_{\rm flip}\ket{2j+1}=\ket{2j}$.

For $d=2$, the four coin states are paired as $\ket{0}(=\ket{00}_2)\leftrightarrow\ket{1} (=\ket{01}_2)$ and $\ket{2} (=\ket{10}_2)\leftrightarrow\ket{3} (=\ket{11}_2)$, corresponding to the opposite directions along the $x$ and $y$ axes, respectively.  In both pairs, the direction reversal is implemented by flipping the least significant coin qubit.  

For $d=3$, the six coin states are paired as $\ket{0}\leftrightarrow\ket{1}$, $\ket{2}\leftrightarrow\ket{3}$, and $\ket{4}\leftrightarrow\ket{5}$, corresponding to the opposite directions along the $x$, $y$, and $z$ axes, respectively.  Since the coin register requires $n_l=3$ qubits, it also contains two unused states, $\ket{6}$ and $\ket{7}$, whose amplitudes are initialized to zero.  The $\hat{X}$ gate on the least significant coin qubit exchanges these two states without mixing them with the six valid coin states.  In general, since $k=2d$ is even, the $\hat{X}$ gate on the least significant coin qubit pairs the unused states $\{\ket{2d},\ldots,\ket{2^{n_l}-1}\}$ only among themselves. Thus, the same direction-reversal circuit applies to any $d$.

\subsection{Non-periodic boundary condition}
The coin-state encoding introduced in the previous section also applies to non-periodic boundary conditions.  In this case, the walker moves only if the neighboring site $\mathbf{x}+\boldsymbol{\delta}_l$ lies within the lattice. Otherwise, its position and coin state remain unchanged; that is, the transition across the boundary is replaced by a self-loop. We use the terms non-periodic and open boundary conditions interchangeably.

To implement this position-dependent shift, we introduce an auxiliary qubit initialized to $\ket{0}_a$, as shown in Fig.~\ref{fig:proposed_circuit}(b).  We define the time-evolution quantum state as
\begin{equation}
    \ket{\Psi(t)}\ket{0}_a
    =
    [\hat{W}_{\rm np}\hat{R}]^t
    \ket{\Psi(0)}\ket{0}_a,
    \label{eq:non-periodic-circuit}
\end{equation}
where $\hat{R}$ is oracle operator same as previous section.  The walk operator is
\begin{equation}
    \hat{W}_{\rm np}
    =\hat{U}_a\,c\text{-}\hat{S}\,\hat{U}_a\hat{C},
    \label{eq:Wnp}
\end{equation}
where $\hat{U}_a$ is a boundary-check operator and $c\text{-}\hat{S}$ is a controlled flip-flop shift. As shown below, the auxiliary qubit returns to $\ket{0}_a$ after each step, so that the success probability is given by Eq.~\eqref{eq:time_prob}.

The operator $\hat{U}_a$ evaluates the move-validity flag $a(\mathbf{x},l)$, defined below, into the auxiliary qubit:
\begin{equation}
    \hat{U}_a
    \ket{\mathbf{x}}\ket{l}\ket{b}_a
    =
    \ket{\mathbf{x}}\ket{l}\ket{b\oplus a(\mathbf{x},l)}_a,
    \label{eq:Ua}
\end{equation}
where $b\in\{0,1\}$ labels the computational basis state of the auxiliary qubit and $\oplus$ denotes addition modulo 2. Starting from $\ket{0}_a$, the auxiliary qubit is thus marked as $\ket{1}_a$ only for allowed moves. Here, the move-validity flag is defined as
\begin{equation}
    a(\mathbf{x},l)=
    \begin{cases}
    1, & \mathbf{x}+\boldsymbol{\delta}_l\in V,\\
    0, & \text{otherwise},
    \end{cases}
    \label{eq:validity-flag}
\end{equation}
where $\mathbf{x}+\boldsymbol{\delta}_l$ is evaluated without the modulo-$L$ reduction used in Eq.~\eqref{eq:flip-flop_walk}. For a $d=2$ hypercubic lattice with $\mathbf{x}=(x,y)$, the flag is given by
\begin{equation}
    a(\mathbf{x},l)=
    \begin{cases}
    1, & x>0,\ l=0,\\
    1, & x<L-1,\ l=1,\\
    1, & y>0,\ l=2,\\
    1, & y<L-1,\ l=3,\\
    0, & \text{otherwise}.
    \end{cases}
    \label{eq:validity-flag-2d}
\end{equation}
As shown in the dotted box of Fig.~\ref{fig:proposed_circuit}(b), $\hat{U}_a$ is implemented by comparators on the position registers, each controlled by the corresponding coin value and flipping the auxiliary qubit when the corresponding condition in Eq.~\eqref{eq:validity-flag-2d} is satisfied. One auxiliary qubit is sufficient to represent this flag, although additional ancilla qubits may be required during circuit synthesis.

The controlled shift applies the flip-flop shift conditioned on the marked auxiliary qubit,
\begin{equation}
    c\text{-}\hat{S}
    =\hat{S}\otimes\ket{1}\bra{1}_a
    +\hat{I}\otimes\ket{0}\bra{0}_a,
    \label{eq:cS}
\end{equation}
where $\hat{S}$ is the flip-flop shift for the periodic case defined in Eq.~\eqref{eq:flip-flop_walk}.  The second $\hat{U}_a$ in Eq.~\eqref{eq:Wnp} then uncomputes the flag, giving
\begin{equation}
    \hat{U}_a\,c\text{-}\hat{S}\,\hat{U}_a
    \ket{\mathbf{x}}\ket{l}\ket{0}_a
    =
    \begin{cases}
    \ket{\mathbf{x}+\boldsymbol{\delta}_l}\ket{\bar{l}}\ket{0}_a, & a(\mathbf{x},l)=1,\\
    \ket{\mathbf{x}}\ket{l}\ket{0}_a, & a(\mathbf{x},l)=0.
    \end{cases}
    \label{eq:uncompute}
\end{equation}
For $a(\mathbf{x},l)=1$, the flag is reset because the reverse move is also allowed, $a(\mathbf{x}+\boldsymbol{\delta}_l,\bar{l})=1$. For $a(\mathbf{x},l)=0$, the state is unchanged, which realizes the self-loop at the boundary.

\section{Simulations}
\label{sec:experiment}
\begin{figure}[ht]
    \centering
    \includegraphics[width=1.0\linewidth]{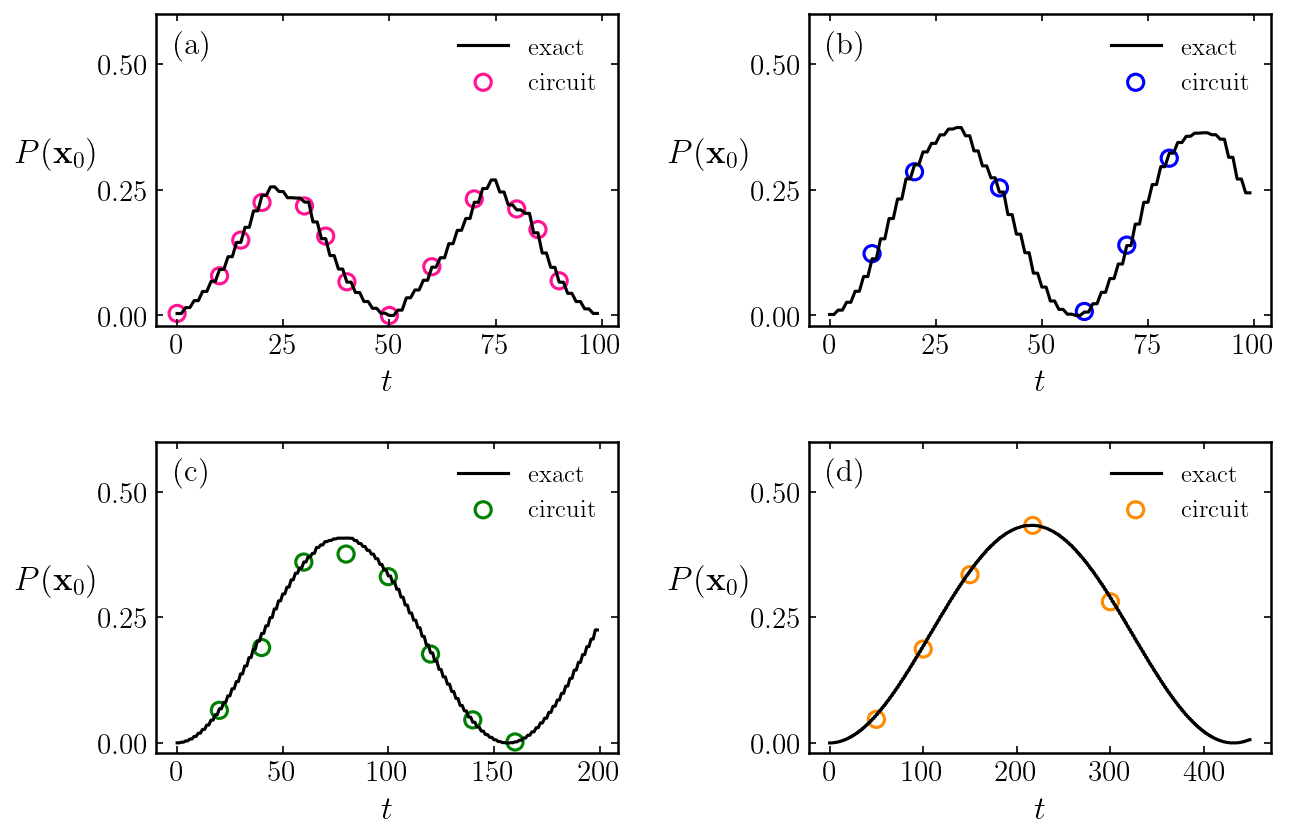}
    \caption{
    The success probability $P(\mathbf{x}_0)$ of finding the target site $\mathbf{x}_0$ as a function of the number of steps $t$ for the quantum spatial search on $d$-dimensional hypercubic lattices with periodic boundary conditions.  The solid lines show the exact numerical results, and open circles show the circuit simulation results obtained with $1000$ shots. (a) $d=2$, $L=16$ and $N=256$. (b) $d=3$, $L=8$ and $N=512$. (c) $d=4$, $L=8$ and $N=4096$. (d) $d=5$, $L=8$ and $N=32768$.}
    \label{fig:ss_search_eval}
\end{figure}

\begin{figure}[ht]
    \centering
    \includegraphics[width=0.8\linewidth]{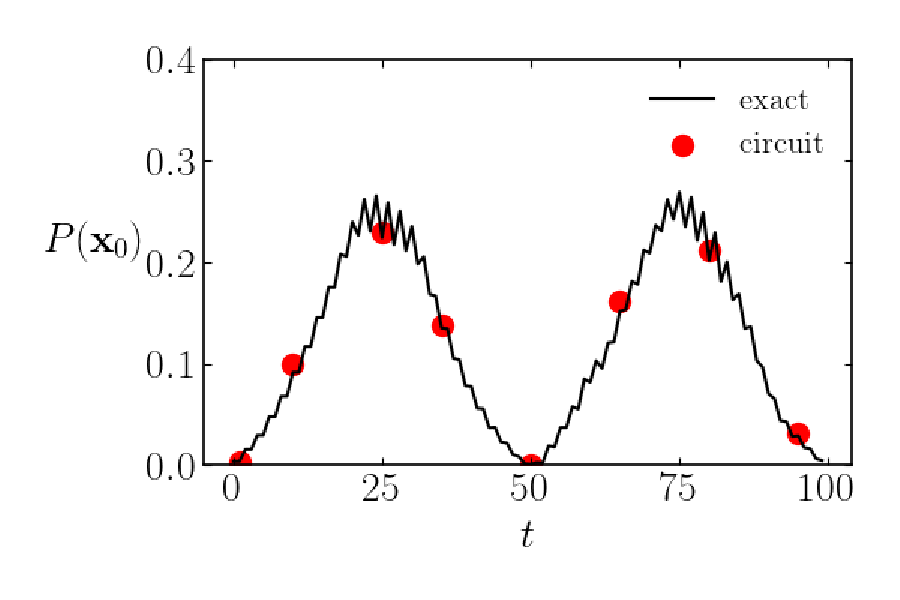}
    \caption{Time evolution of the success probability $P({\bf x}_0)$ under non-periodic boundary conditions on a two-dimensional lattice with $N=256$.}
    \label{fig:ss_non_periodic}
\end{figure}

\begin{figure*}[ht]
    \centering
    \includegraphics[width=0.8\linewidth]{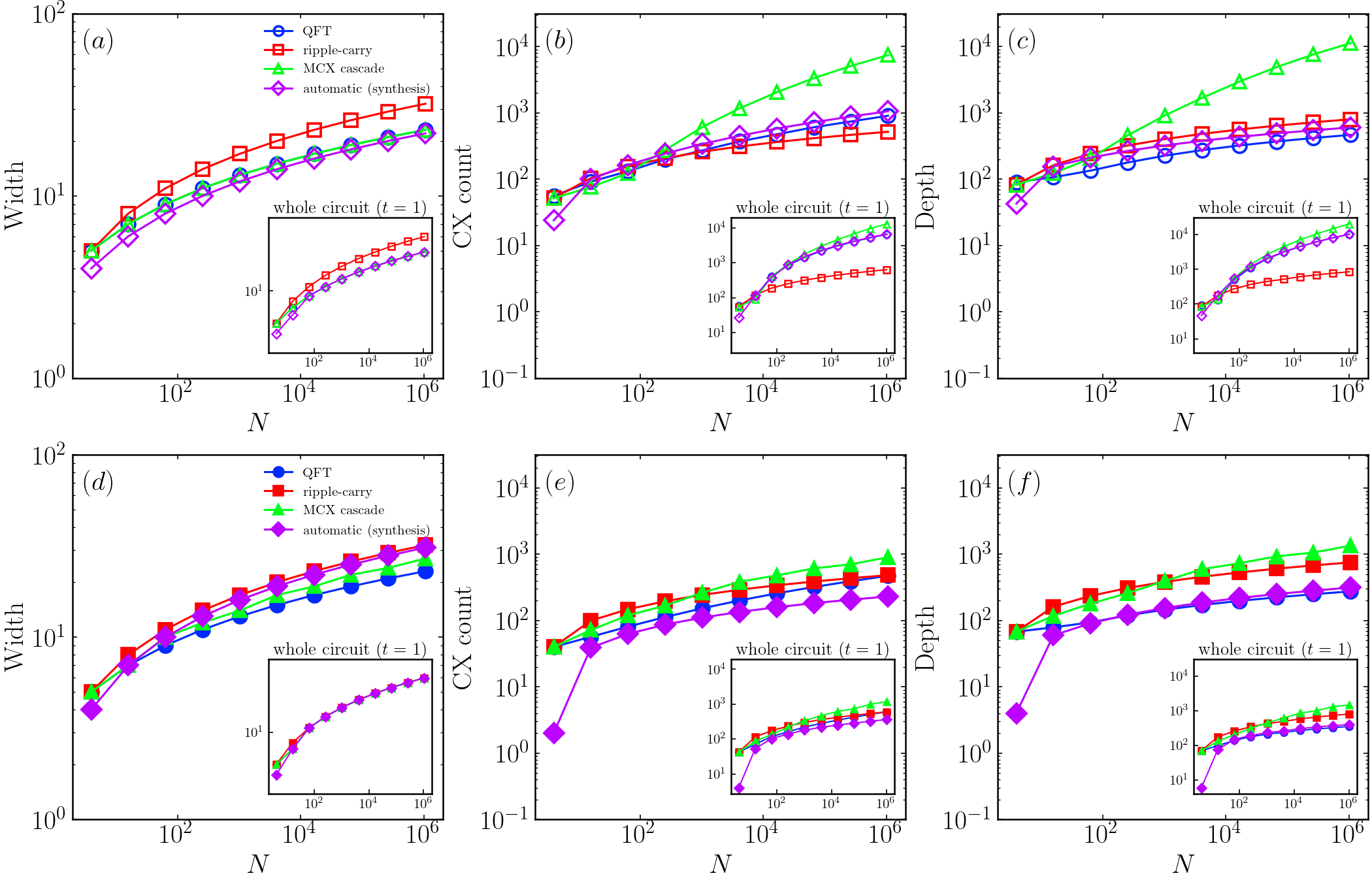}
    \caption{
    Circuit resources of the shift operator $\hat S_{\rm move}$ on a two-dimensional periodic lattice with $N=L^2$ sites, where $L=2^n$ ($n=1,\ldots,10$). The columns show circuit width (left), CX count (middle), and depth (right).  (a)--(c) show default synthesis (open markers), while (d)--(f) show CX-optimized synthesis with basis $\{\mathrm{CX},U\}$ (filled markers). Colors and marker shapes distinguish four implementations: QFT-based, ripple-carry, MCX-based, and automatically synthesized arithmetic. The main panels show a single application of $\hat S_{\rm move}$, while the insets show the complete search circuit for one iteration ($t=1$), including initial-state preparation, the oracle, and the coin operator.
    }
    \label{fig:comparison-shift-operators}
\end{figure*}

\begin{figure}[ht]
    \centering
    \includegraphics[width=1.0\linewidth]{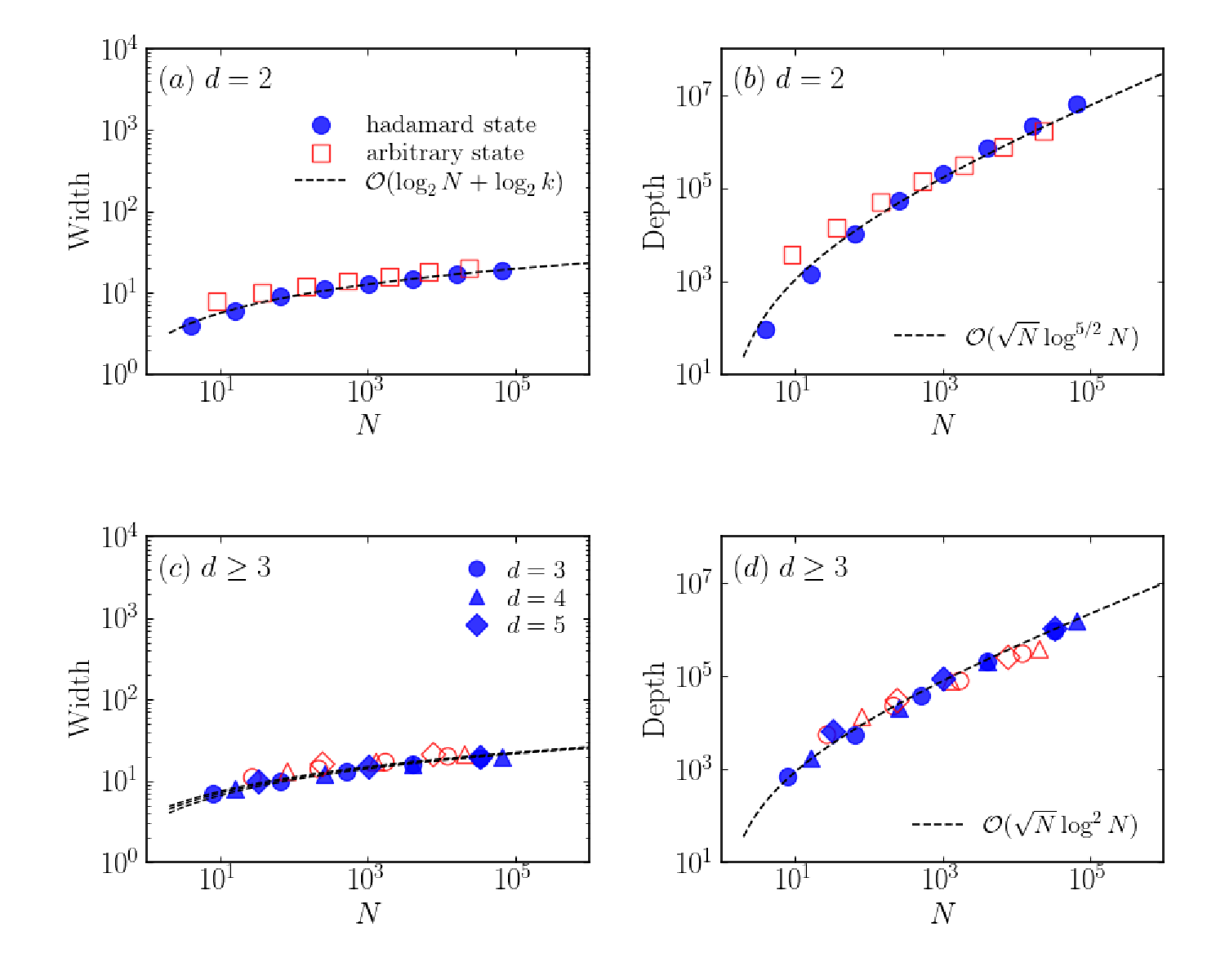}
    \caption{
    Quantum-resource scaling with problem size $N$ for spatial search in $d=2,3,4$, and $5$.  (a),(b) show circuit width and depth for $d=2$, while (c),(d) show the corresponding results for $d=3,4$, and $5$. Filled blue and open red markers represent Hadamard and arbitrary initial states, respectively. In (c),(d), circles, triangles, and diamonds correspond to $d=3,4$, and $5$.  Only the prefactors are fitted to the Hadamard data. The number of iterations is set to $t=\sqrt{N\log N}$ for $d=2$ and $t=\sqrt{N}$ for $d\geq3$.  The circuits are synthesized using the automatic shift implementation.}
    \label{fig:resources_2d3d}
\end{figure}

\begin{figure}[ht]
    \centering
    \includegraphics[width=1.0\linewidth]{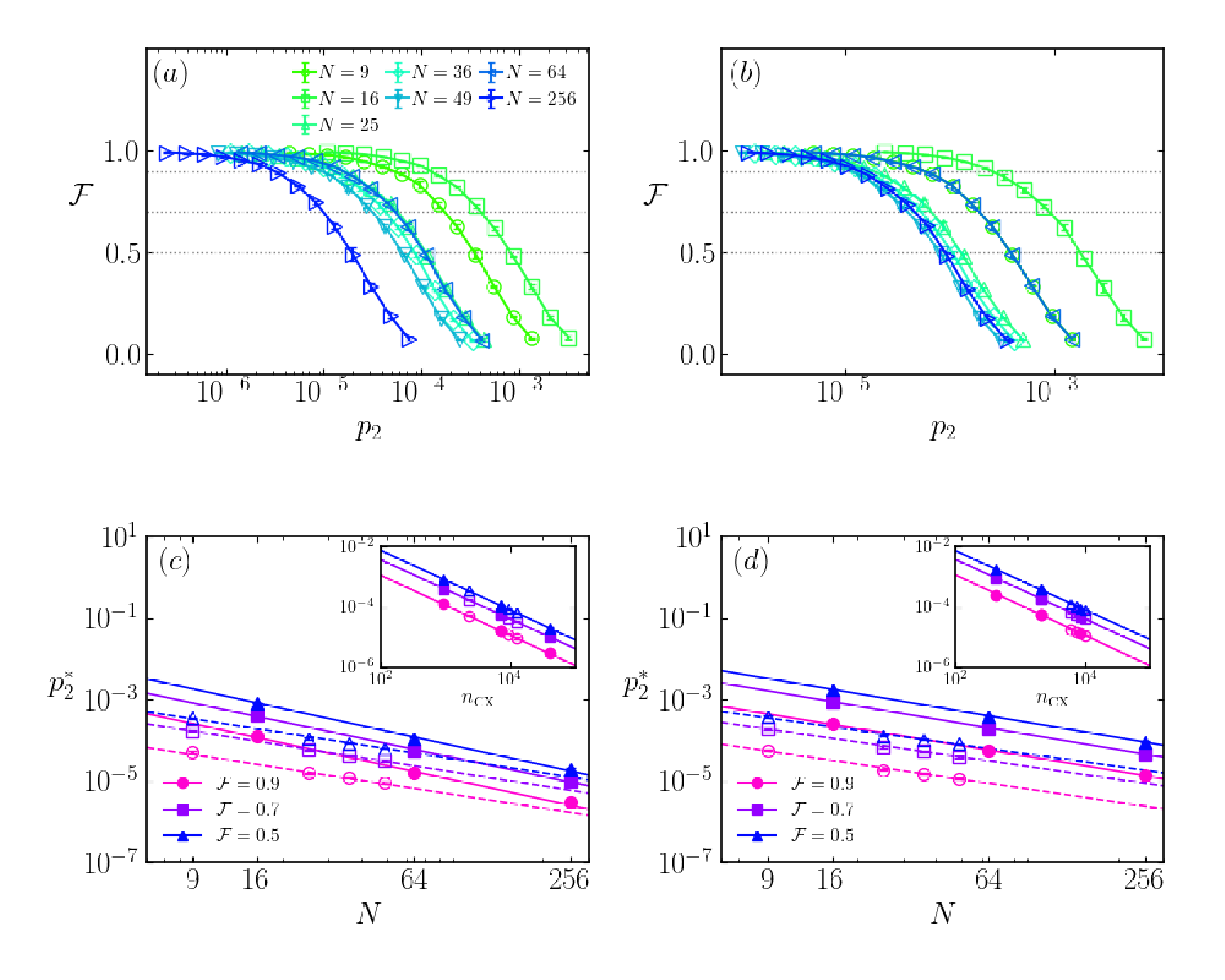}
    \caption{Fidelity scaling of quantum spatial search on two-dimensional lattices under CX-gate depolarizing noise, without (left) and with (right) CX-count optimization.  (a) and (b) show fidelity $\mathcal{F}$ versus two-qubit error rate $p_2$ for different lattice sizes $N$, with dotted lines indicating $\mathcal{F}=0.9,0.7$, and $0.5$. (c) and (d) show the threshold error rate $p_2^*$, defined as the value of $p_2$ at which the fidelity reaches each of these thresholds, as a function of $N$. Solid and dashed lines show weighted power-law fits for $L=2^{n_x}$ and $L\neq2^{n_x}$, respectively. Insets show $p_2^*$ versus the total CX count $n_{\mathrm{CX}}$.}
    \label{fig:fidelity-scaling}
\end{figure}

\begin{table}[ht]
    \centering
    \caption{Fitted scaling exponents of the threshold error rate under CX-gate depolarizing noise, defined by $p_2^*\propto N^\alpha$ and $p_2^*\propto n_{\mathrm{CX}}^\beta$. Numbers in parentheses indicate one standard deviation in the last digit.}
    \label{tab:fidelity-scaling}
    \begin{ruledtabular}
    \begin{tabular}{ccccc}
        \multirow{2}{*}{Circuit}
        & \multirow{2}{*}{$\mathcal{F}$}
        & \multicolumn{2}{c}{$\alpha$}
        & \multirow{2}{*}{$\beta$} \\
        \cline{3-4}
        & & $L=2^{n_x}$ & $L\neq 2^{n_x}$ & \\
        \hline
        & $0.5$ & $-1.39(5)$ & $-0.98(4)$ & $-0.991(1)$ \\
        Default & $0.7$ & $-1.34(7)$ & $-1.00(4)$ & $-0.984(5)$ \\
        & $0.9$ & $-1.3(1)$  & $-0.9(1)$  & $-0.99(2)$ \\
        \hline
        & $0.5$ & $-1.07(1)$ & $-0.88(8)$ & $-0.982(3)$ \\
        CX-opt.& $0.7$ & $-1.07(2)$ & $-0.91(4)$ & $-0.989(3)$ \\
        & $0.9$ & $-1.04(2)$ & $-0.93(7)$ & $-1.00(1)$ \\
    \end{tabular}
    \end{ruledtabular}
\end{table}

In this section, we first validate the correctness of the constructed quantum circuits. We then compare the circuit resources required by different implementations of the shift operator and evaluate the resource scaling of quantum spatial search circuits on hypercubic lattices with $d=2,3,4$, and $5$ using the Classiq synthesis engine~\cite{goldfriend2024design}.  Finally, we evaluate the fidelity between the ideal and noisy states under a CX-gate depolarizing noise model.  

We first validate the constructed circuits by comparing the simulation results with exact theoretical calculations. Figure~\ref{fig:ss_search_eval} shows the time evolution of the success probability at the target site on periodic hypercubic lattices with $d=2,3,4$, and $5$. The solid lines represent the exact theoretical results, while the circles show the circuit simulation results obtained using Qmod~\cite{vax2025qmod} and the Classiq synthesis engine~\cite{goldfriend2024design}.  For each time step, the success probability is estimated from $1000$ measurement shots.  The circuit simulation results agree well with the theoretical values for all dimensions considered.

We also validate the circuit construction for $d=2$ under non-periodic boundary conditions.  Figure~\ref{fig:ss_non_periodic} shows the time evolution of the success probability at the marked site. The circuit simulation results agree well with the exact theoretical calculations, confirming the correctness of the proposed construction.

The shift operator can be implemented using arithmetic operations on the position registers, including quantum Fourier transform (QFT)-based adders~\cite{shakeel2020efficient,draper2000addition,koch2022gate}, MCX gates~\cite{puengtambol2021implementation,Sahu2024,douglas2009efficient}, and ripple-carry adders~\cite{cuccaro2004new}.  A recent study~\cite{febrianto2026quantum} compared MCX-based and ripple-carry-based shift circuits, showing that the latter requires fewer gates.  The MCX-based implementation considered below follows the construction used in previous quantum-walk circuits~\cite{Sahu2024,douglas2009efficient} and thus serves as a direct benchmark against existing shift-operator implementations.  However, systematic comparisons including QFT-based implementations and their impact on the overall quantum spatial search circuit remain limited. We therefore compare these approaches in terms of circuit width, CX-gate count, and circuit depth.

Figure~\ref{fig:comparison-shift-operators} compares the circuit resources of four implementations of the shift operator $\hat{S}_{\rm move}$ on a $d=2$ lattice: QFT-based, ripple-carry, MCX-based, and automatically synthesized arithmetic implementations.  In the last approach, the modular addition is described as a high-level arithmetic operation in Qmod, and its gate-level implementation is generated during synthesis under the specified optimization setting.  Figs.~\ref{fig:comparison-shift-operators}(a)--(c) show the results without an optimization objective (default), while Figs.~\ref{fig:comparison-shift-operators}(d)--(f) show those with CX-gate count minimization (CX-opt).  The main panels show the resources for a single application of $\hat{S}_{\rm move}$, while the insets show those for one search iteration ($t=1$), including initial-state preparation, the oracle, and the coin operator, as illustrated in Fig.~\ref{fig:proposed_circuit}(a).  Circuit simulations also confirm agreement with the exact theoretical results for the QFT-based, ripple-carry, and MCX-based implementations (see Appendix~\ref{appdx:shift} and Fig.~\ref{fig:success-prob-different-circuit}).

We first compare the circuit resources for a single application of $\hat{S}_{\rm move}$. The MCX-based implementation has the largest CX-gate count and circuit depth under both optimization settings, with the fastest growth as the position-register size $n_x$ increases.

In the default setting, the ripple-carry implementation achieves the lowest CX-gate count, while the QFT-based, ripple-carry, and automatically synthesized arithmetic implementations show similar scaling in circuit depth.

Under CX-opt, the ripple-carry and automatically synthesized arithmetic implementations exhibit similar scaling in CX-gate count, with the latter achieving a lower count through the use of additional ancilla qubits. The QFT-based implementation achieves a slightly lower circuit depth for larger $N$, while the other two implementations show similar depth scaling. These differences suggest that the use of ancilla qubits enables more efficient gate arrangements, affecting the prefactors rather than the overall scaling behavior.  The detailed circuit constructions for the QFT-based, ripple-carry, and MCX-based implementations are provided in Appendix~\ref{appdx:shift}.

We next compare the circuit resources for one complete search iteration, as shown in the insets of Fig.~\ref{fig:comparison-shift-operators}. In the default setting, the multi-controlled phase flip of the oracle dominates the circuit resources, resulting in little difference between the QFT-based and automatically synthesized arithmetic implementations. In contrast, the ripple-carry implementation reduces the CX-gate count by approximately one order of magnitude. This reduction is attributed mainly to the reuse of ancilla qubits from the shift operator in the oracle, rather than to the efficiency of the shift operator alone.

Under CX-opt, the four implementations have comparable circuit widths. The automatically synthesized arithmetic implementation achieves the lowest CX-gate count, while maintaining a circuit depth comparable to that of the QFT-based implementation.

These results highlight the trade-offs among circuit width, CX-gate count, and circuit depth. The MCX-based implementation offers no resource advantage under the conditions considered here. Furthermore, the resources of the complete search circuit depend not only on the shift implementation but also on the oracle and the availability of ancilla qubits.

We further evaluate the scaling of circuit width and depth for quantum spatial search on $d$-dimensional hypercubic lattices.  We set the number of search steps to $t_{\rm opt}=\sqrt{N\log N}$ for $d=2$ and $t_{\rm opt}=\sqrt{N}$ for $d\geq3$, and synthesize circuits containing all $t_{\rm opt}$ iterations.  Here, $t_{\rm opt}$ denotes the optimal number of steps at which the success probability reaches its first peak.  We set $t_{\rm opt}$ according to its asymptotic scaling without the constant prefactor, so it does not necessarily coincide exactly with the first peak in Fig.~\ref{fig:ss_search_eval}.  

Figure~\ref{fig:resources_2d3d} shows the resulting circuit resources for $d=2,3,4$, and $5$. The dashed lines indicate the expected asymptotic scaling: $\mathcal{O}(\log N)$ for circuit width, $\mathcal{O}(\sqrt{N}(\log N)^{5/2})$ for circuit depth at $d=2$, and $\mathcal{O}(\sqrt{N}(\log N)^2)$ for circuit depth at $d\geq3$. The numerical results are consistent with these asymptotic upper bounds over the investigated range of lattice sizes.

The depth scaling observed in Fig.~\ref{fig:resources_2d3d} can be understood from the circuit structure.  The total circuit depth is expressed as
\begin{equation}
    D = D_{\rm init} + t_{\rm opt}D_{\rm step},
    \label{eq:depth-of-circuit}
\end{equation}
where $D_{\rm init}$ is the initial-state preparation depth, $t_{\rm opt}$ is the number of search steps, and $D_{\rm step}$ is the depth of one search step.  Each search step consists of the oracle, coin, and shift operators, giving $D_{\rm step}=D_R+D_C+D_S$, where $D_R$, $D_C$, and $D_S$ denote their respective depths.

The state preparation of a general $n$-qubit state requires at most $\mathcal{O}(2^n)$ circuit depth~\cite{10.1145/3719276.3725183}. Assuming independent preparation of each position register, the initialization depth is bounded by $D_{\rm init}=\mathcal{O}(2^{n_x})=\mathcal{O}(N^{1/d})$ for fixed $d$. For Hadamard-based initialization, this reduces to $D_{\rm init}=\mathcal{O}(1)$.

The ancilla-free decomposition of the multi-controlled oracle gives a depth upper bound of $D_R=\mathcal{O}(\log^2 N)$~\cite{PhysRevA.52.3457}. For fixed $d$, the coin operator acts on a constant number of qubits, giving $D_C=\mathcal{O}(1)$.

For the shift operator, we consider the ripple-carry implementation, which requires a circuit depth of $\mathcal{O}(n_x)$ for each increment or decrement operation based on the result of Fig.~\ref{fig:comparison-shift-operators}. As shown in Fig.~\ref{fig:proposed_circuit}(c), the shift operator performs coin-controlled increments and decrements along each of the $d$ spatial axes, requiring $2d$ conditional operations. Assuming sequential execution, the total shift depth is bounded by $D_S=\mathcal{O}(2dn_x)=\mathcal{O}(\log N)$ for fixed $d$. Combining this with the oracle and coin operators, the depth of one search step is bounded by
\begin{align}
    D_{\rm step}
    &= D_R+D_C+D_S \notag\\
    &= \mathcal{O}(\log^2 N)
     + \mathcal{O}(1)
     + \mathcal{O}(\log N) \notag\\
    &= \mathcal{O}(\log^2 N).
\end{align}

Therefore, for $d=2$, the total circuit depth is bounded by
\begin{align}
    D_{d=2}
    &= \mathcal{O}(N^{1/2})
    + \mathcal{O}(\sqrt{N\log N})\mathcal{O}(\log^2 N)
    \notag\\
    &= \mathcal{O}(\sqrt{N}(\log N)^{5/2}).
    \label{eq:scaling-d2}
\end{align}
Similarly, for $d\geq3$,
\begin{align}
    D_{d\geq3}
    &= \mathcal{O}(N^{1/d})
    + \mathcal{O}(\sqrt{N})\mathcal{O}(\log^2 N)
    \notag\\
    &= \mathcal{O}(\sqrt{N}\log^2 N).
    \label{eq:scaling-d3}
\end{align}
These results establish asymptotic upper bounds on the total circuit depth for fixed $d$.  The scaling is determined by the number of search steps and the depth of each step, which is dominated by the $\mathcal{O}(\log^2 N)$ contribution from the oracle.

Finally, we evaluate the robustness of quantum spatial search circuits against CX-gate depolarizing noise.  For $d=2$, circuits with $t=\sqrt{N\log N}$ search steps are synthesized using all-to-all connectivity.  We compare circuits synthesized with the default settings and with CX-count optimization.

The depolarizing channel is defined as~\cite{Nielsen_Chuang_2010}
\begin{equation}
    \mathcal{E}(\rho)
    =p\frac{\hat{I}}{2^n}+(1-p)\rho,
\end{equation}
where $p$ is the depolarizing error rate and $n$ is the number of qubits affected by the channel. We apply the two-qubit channel ($n=2$) with error rate $p_2$ after each CX gate, while neglecting all other error sources.

We quantify the effect of noise using the fidelity between the ideal state $\ket{\Psi_{\rm ideal}}$ and the noisy state $\rho_{\rm noisy}$,
\begin{equation}
    \mathcal{F}
    =\langle\Psi_{\rm ideal}|\rho_{\rm noisy}|\Psi_{\rm ideal}\rangle,
    \label{eq:fidelity}
\end{equation}
which is estimated by the quantum-trajectory method~\cite{plenio1998quantum} using the statevector method of the Qiskit Aer simulator~\cite{qiskit2024} as $\mathcal{F}\simeq \frac{1}{R}\sum_{r=1}^{R}|\langle\Psi_{\rm ideal}|\phi_r\rangle|^2$, where $\ket{\phi_{r}}$ is the final state of the $r$-th noisy trajectory and $R=10^3$.  Each data point represents the mean of three independent runs, with error bars indicating the standard deviation across runs. 

Figure~\ref{fig:fidelity-scaling} shows the results for $d=2$ hypercubic lattices with $N=9,16,25,36,49,64,256$, where the left and right columns correspond to the default and CX-optimized circuits, respectively. As shown in Figs.~\ref{fig:fidelity-scaling}(a) and (b), the fidelity decreases with $p_2$, and larger $N$ requires lower error rates to maintain the same fidelity.

To quantify this behavior, we define the threshold error rate $p_2^*$ as the value of $p_2$ at which the fidelity reaches $\mathcal{F}=0.9$, $0.7$, or $0.5$. Figs.~\ref{fig:fidelity-scaling}(c) and (d) show $p_2^*$ as a function of $N$.  We fit $p_2^*\propto N^\alpha$ separately for $L=2^{n_x}$ and $L\neq2^{n_x}$, since the latter requires general initial-state preparation instead of Hadamard gates, as well as modular arithmetic modulo $L$ to implement the periodic boundary conditions.  As summarized in Table~\ref{tab:fidelity-scaling}, CX optimization changes $\alpha$ from approximately $-1.4$ to $-1.06$ for $L=2^{n_x}$, indicating a slower decrease in $p_2^*$ with increasing $N$.  For $L\neq2^{n_x}$, a smaller improvement is observed, with $\alpha$ changing from approximately $-1.0$ to $-0.9$. 

The insets of Figs.~\ref{fig:fidelity-scaling}(c) and (d) show $p_2^*$ as a function of the total CX-gate count $n_{\mathrm{CX}}$.  All data follow $p_2^*\propto n_{\mathrm{CX}}^\beta$ with $\beta\simeq-1$ shown in Table~\ref{tab:fidelity-scaling}.  This behavior can be explained by a simple independent-error model, in which the circuit fidelity is approximated by the product of individual gate fidelities~\cite{boixo2018characterizing}.  Considering only CX-gate errors, we obtain
\begin{equation}
    \mathcal{F}\sim(1-p_2)^{n_{\mathrm{CX}}}
    \simeq e^{-p_2n_{\mathrm{CX}}},
\end{equation}
for $p_2\ll1$.  For a fixed target fidelity $\mathcal{F}^*$, this gives $p_2^*\simeq-\ln\mathcal{F}^*/n_{\mathrm{CX}}\propto n_{\mathrm{CX}}^{-1}$, consistent with the observed scaling $\beta\simeq-1$.

We also evaluate robustness against initial-state preparation errors and spatial asymmetry in Appendices~\ref{appdx:robut-init} and~\ref{appdx:robut-spatial-asymmetry}.

\section{Discussion}
\label{sec:discussion}

The circuit-resource analysis shows that the circuit width is determined primarily by the position and coin registers, requiring approximately $\log_2 N+\lceil\log_2 k\rceil$ qubits, where $N$ is the number of lattice vertices and $k=2d$ is the coin-space dimension. Thus, for fixed $d$, the circuit width grows logarithmically with the search-space size.

The total circuit depth depends on the number of search steps and the depth of each step. The number of search steps scales as $\mathcal{O}(\sqrt{N\log N})$ for $d=2$ and $\mathcal{O}(\sqrt{N})$ for $d\geq3$. For the circuit decomposition considered here, the depth of each search step is dominated by the oracle, which has an upper bound of $\mathcal{O}(\log^2 N)$.   Combining these contributions gives asymptotic depth upper bounds of $\mathcal{O}(\sqrt{N}\log^{5/2} N)$ for $d=2$ and $\mathcal{O}(\sqrt{N}\log^2 N)$ for $d\geq3$, consistent with the numerical results. While a previous study~\cite{Sahu2024} estimated the CNOT-gate complexity of two-dimensional quantum-walk search primarily from the shift operator, our work provides a systematic analysis of the total circuit depth, including the oracle, coin, and shift operators, for $d$-dimensional hypercubic lattices.

The circuit widths and depths are nearly identical for Hadamard-based and general initial-state preparation over the investigated range. Although general state preparation requires a depth of $\mathcal{O}(N^{1/d})$ under the considered construction, it is performed only once and remains subleading compared with the repeated search steps. Consequently, the total circuit depth is primarily determined by the search operations rather than by initial-state preparation.

The shift-operator comparison in Fig.~\ref{fig:comparison-shift-operators} reveals tradeoffs among the implementations. The MCX-based implementation becomes increasingly costly as the register size grows. The QFT-based implementation achieves the smallest circuit width and depth, but requires $\mathcal{O}(n_x^2)$ controlled-phase gates. In contrast, the ripple-carry implementation has a CX-gate count that scales linearly with $n_x$, at the cost of additional ancilla qubits. With CX-count optimization, the automatically synthesized implementation achieves the lowest CX count among the investigated methods.  Its circuit depth is also comparable to that of the QFT-based implementation, although it requires a few additional ancilla qubits. These results demonstrate that the preferred shift implementation depends on whether circuit width, depth, or CX-gate count is prioritized.

The noise analysis further shows that larger search spaces require lower CX-gate error rates to maintain the same fidelity. For $L=2^{n_x}$, CX optimization changes the fitted exponent $\alpha$ in $p_2^*\propto N^\alpha$ from approximately $-1.4$ to $-1.06$, indicating a slower decrease in the tolerable error rate. For $L\neq2^{n_x}$, a smaller improvement is observed, with $\alpha$ increasing from approximately $-1.0$ to $-0.9$.

The observed noise scaling is consistent with previous studies of gate-error accumulation. The effective fidelity can decay exponentially with the number of two-qubit gates contributing to an observable~\cite{kechedzhi2024effective}, and gate-level noise has also been shown to degrade quantum-walk search performance~\cite{mammola2026noisy}. In our simulations, the threshold error rate follows $p_2^*\propto n_{\mathrm{CX}}^\beta$ with $\beta\simeq-1$, consistent with the independent-error approximation $\mathcal{F}\sim e^{-p_2n_{\mathrm{CX}}}$. This indicates that reducing the total CX-gate count improves noise tolerance under the considered noise model.

The present noise analysis is limited to CX-gate depolarizing errors under all-to-all connectivity and problem sizes up to $N=256$. It does not account for hardware connectivity, routing overhead, calibration-dependent errors, or other gate and measurement errors. The observed scaling $p_2^*\propto n_{\mathrm{CX}}^{-1}$ should therefore be interpreted within these limitations.

Although CX-count optimization improves noise tolerance, the threshold error rate still decreases with increasing problem size. Scaling to larger systems will therefore require further circuit optimization, lower physical gate errors, or fault-tolerant implementations. Future work should investigate the effects of realistic device connectivity and noise models on circuit fidelity and search success probability, as well as evaluate the circuits on quantum hardware.

Another direction is to extend the circuit construction to irregular structures, such as defective and fractal lattices~\cite{daSTeles2021,PhysRevA.87.012314,PhysRevA.86.012332,PhysRevA.101.022312}.  Our implementation of non-periodic boundary conditions uses position-dependent controls to prevent moves beyond the lattice boundaries. A similar approach could be applied to defective lattices by preventing moves to missing sites.  For fractal lattices such as the Sierpinski carpet, however, checking individual missing sites may become costly as the lattice size increases.  Developing efficient validity-check circuits that exploit the self-similar structure of fractal lattices is therefore an important direction for future work.

\section{Conclusion}
\label{sec:conclusion}
In this study, we developed quantum circuits for spatial search based on discrete-time quantum walks on $d$-dimensional hypercubic lattices.  The proposed circuit implements the flip-flop shift using coin-controlled position updates and a single Pauli-$X$ gate to reverse the walker direction. Numerical simulations confirmed that the circuits reproduce the theoretical time evolution of quantum spatial search for $d=2,3,4$, and $5$.  We also verified the construction under non-periodic boundary conditions for $d=2$.

We analyzed the circuit resources and showed that the circuit width scales as $\mathcal{O}(\log N)$ for fixed $d$. We derived asymptotic upper bounds on the total circuit depth of $\mathcal{O}(\sqrt{N}\log^{5/2} N)$ for $d=2$ and $\mathcal{O}(\sqrt{N}\log^2 N)$ for $d\geq3$, consistent with the numerical results. A comparison of QFT-based, ripple-carry, MCX-based, and automatically synthesized shift implementations revealed trade-offs among circuit width, depth, and CX-gate count.  The ripple-carry and automatically synthesized implementations, which use additional ancilla qubits, achieved lower CX counts, whereas the QFT-based implementation achieved the smallest circuit width and depth.

We further evaluated robustness against CX-gate depolarizing noise. For $d=2$, the threshold error rate followed $p_2^*\propto n_{\mathrm{CX}}^{-1}$, indicating that reducing the CX-gate count improves noise tolerance under the considered noise model.

\begin{acknowledgments}
The author used ChatGPT (OpenAI, GPT-5 and GPT-6) for language editing and assistance with mathematical verification. The author independently reviewed and verified all scientific content.
\end{acknowledgments}

\section*{Data and Code Availability}
The Qmod source code, synthesis scripts, configuration files, and raw data used to generate the figures in this paper are publicly available at \url{https://github.com/Rei-classiq/quantum_spatial_search_hypercubic_lattice}.

\appendix

\section{Synthesis data}
\label{appdx:synthesis data}

To examine the gate-level cost of a single search iteration, we synthesized the $t=1$ search circuit, which consists of the initial-state preparation followed by one application of the oracle operator $\hat{R}$ and one application of the walk operator $\hat{W}$. For the power-of-two lattices considered here, the position and coin registers are initialized using Hadamard gates applied in parallel, and the initialization depth is therefore unity. All circuits were synthesized using the same backend and synthesis settings as those used for the resource analysis in the main text. Table~\ref{tab:supp_d2_t1} lists the resulting circuit resources.

\begin{table}[ht]
\caption{
Raw synthesis data for one search iteration, $\hat{W}\hat{R}\ket{\Psi(0)}$, on two-dimensional periodic lattices with $L=2^{n_x}$ and $N=L^2$. The initial position and coin states are prepared using Hadamard gates.
}
\label{tab:supp_d2_t1}
\begin{ruledtabular}
\begin{tabular}{rrrrrr}
$L$ & $n_x$ & $N$ & Width & Depth & CX \\
\hline
2    & 1  & 4       & 4  & 47    & 27   \\
4    & 2  & 16      & 6  & 180   & 115  \\
8    & 3  & 64      & 9  & 545   & 377  \\
16   & 4  & 256     & 11 & 1211  & 893  \\
32   & 5  & 1024    & 13 & 2095  & 1501 \\
64   & 6  & 4096    & 15 & 3223  & 2253 \\
128  & 7  & 16384   & 17 & 4593  & 3149 \\
256  & 8  & 65536   & 19 & 6246  & 4189 \\
512  & 9  & 262144  & 21 & 8166  & 5373 \\
1024 & 10 & 1048576 & 23 & 10342 & 6701 \\
\end{tabular}
\end{ruledtabular}
\end{table}

Both the depth and the CX count increase faster than linearly in $n_x$. Since the present analysis focuses on the leading dependence on $n_x$, constant and lower-order terms are omitted, and the data are fitted using quadratic functions without constant or linear terms. The resulting depth fit is
\begin{equation}
    D_{t=1}(n_x)
    \simeq
    99(2)\,n_x^2,
\end{equation}
where the number in parentheses denotes one standard error of the fitted coefficient. The corresponding CX-count fit is
\begin{equation}
    N_{\mathrm{CX},t=1}(n_x)
    \simeq
    66(1)\,n_x^2.
\end{equation}
To check the exponent, we also fit power laws $D_{t=1}\propto n_x^{\gamma}$ and $N_{\mathrm{CX},t=1}\propto n_x^{\gamma}$ to the data with $n_x\geq4$. These fits give $\gamma=2.34(2)$ for the depth and $\gamma=2.19(2)$ for the CX count.  Both exponents approach $2$ when the fitting range is restricted to larger $n_x$ ($\gamma=2.26$ and $2.11$ for $n_x\geq8$), consistent with the expected quadratic scaling.

The $t=1$ circuit includes the unit-depth Hadamard initialization. Since this contribution is independent of $n_x$, it does not affect the leading dependence on the position-register size. The synthesis results are therefore consistent with an effective one-iteration depth of
\begin{equation}
    D_{\mathrm{step}}
    =
    \mathcal{O}(n_x^2)
    =
    \mathcal{O}(\log^2 N).
\end{equation}

Table~\ref{tab:supp_d2_initialization} lists the resources required for general state preparation without applying the oracle or walk operator. The results depend primarily on the allocated register size $2^{n_x}$ rather than on the exact lattice size $L$.

\begin{table}[ht]
\caption{
Raw synthesis data for general initial-state preparation, $\ket{\Psi(0)}$, on two-dimensional lattices with $L\neq2^{n_x}$. No oracle or walk-operator application is included.
}
\label{tab:supp_d2_initialization}
\begin{ruledtabular}
\begin{tabular}{rrrrrr}
$L$ & $n_x$ & $N$ & Width & Depth & CX \\
\hline
3   & 2  & 9      & 8  & 5    & 4    \\
6   & 3  & 36     & 10 & 12   & 12   \\
12  & 4  & 144    & 12 & 21   & 28   \\
23  & 5  & 529    & 14 & 54   & 60   \\
44  & 6  & 1936   & 16 & 99   & 124  \\
82  & 7  & 6724   & 18 & 210  & 252  \\
153 & 8  & 23409  & 20 & 365  & 508  \\
285 & 9  & 81225  & 22 & 868  & 1020 \\
531 & 10 & 281961 & 24 & 1941 & 2044 \\
987 & 10 & 974169 & 24 & 2023 & 2044 \\
\end{tabular}
\end{ruledtabular}
\end{table}

For data points with the same $n_x$, the synthesized resources are nearly identical even when $L$ differs. After averaging the depths obtained for the same $n_x$, the data are fitted as a function of the allocated register dimension $2^{n_x}$. Retaining only the leading dependence gives
\begin{equation}
    D_{\mathrm{init}}^{(\mathrm{arb})}(n_x)
    \simeq
    1.86(5)\,2^{n_x},
\end{equation}
where the number in parentheses denotes one standard error of the fitted coefficient. The CX count follows the exact relation
\begin{equation}
    N_{\mathrm{CX,init}}^{(\mathrm{arb})} = 2^{n_x+1}-4.
\end{equation}
Because $n_x=\lceil\log_2L\rceil$, one has $L\leq2^{n_x}<2L$, and therefore $2^{n_x}=\mathcal{O}(L)=\mathcal{O}(\sqrt{N})$. The general state-preparation depth for the present construction is thus consistent with
\begin{equation}
    D_{\mathrm{init}}^{(\mathrm{arb})}(N)
    =
    \mathcal{O}(\sqrt{N}).
\end{equation}

This general state-preparation cost is incurred only once at the beginning of the circuit. By contrast, the oracle and walk operators are repeatedly applied throughout the search. The $\mathcal{O}(\sqrt{N})$ initialization depth therefore increases the finite-size circuit depth relative to Hadamard-based initialization but does not change the leading $\mathcal{O}(\sqrt{N}\log^{5/2} N)$ depth scaling derived in the main text.

\section{Robustness to Initial State}
\label{appdx:robut-init}
\begin{figure}[ht]
    \centering
    \includegraphics[width=1.0\linewidth]{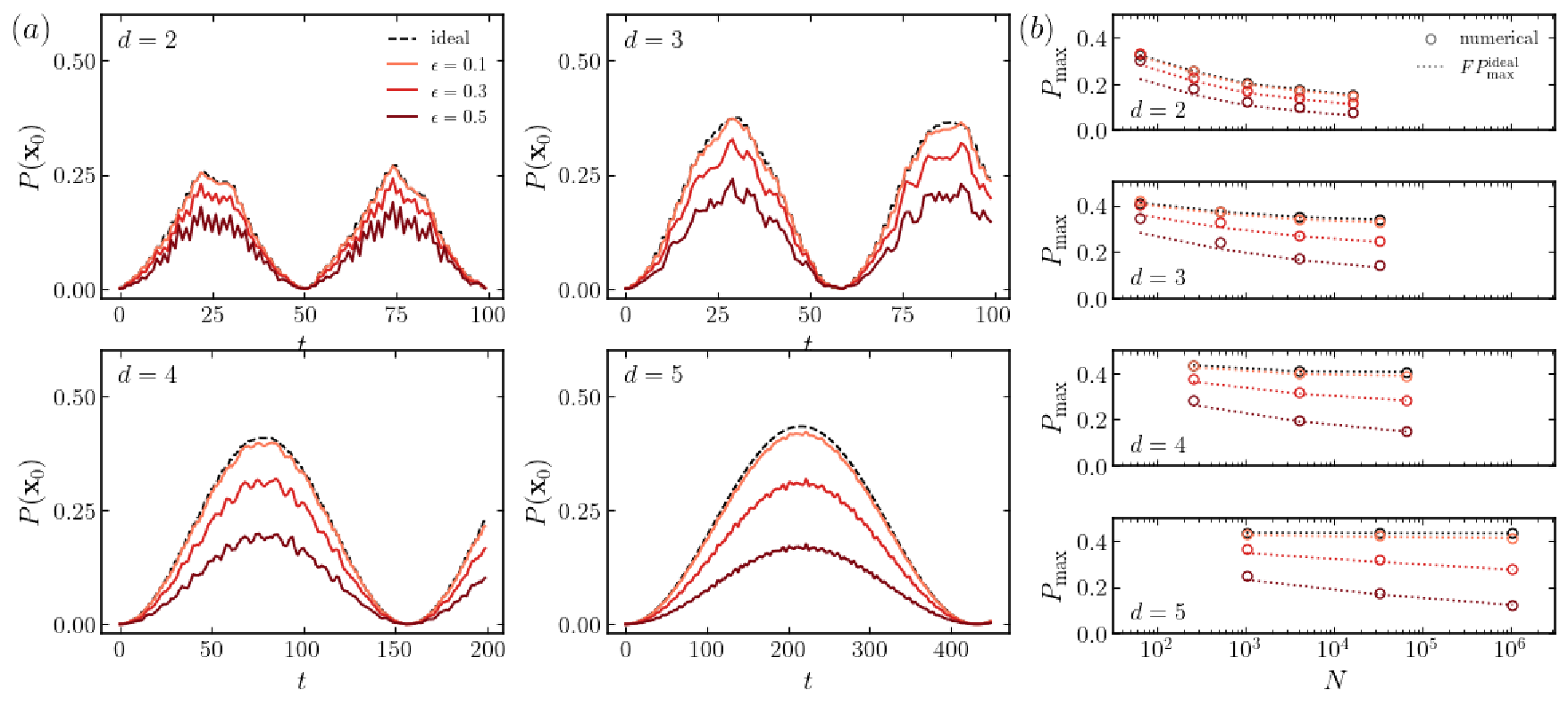}
    \caption{Robustness of spatial search to errors in initial-state preparation. The Hadamard gate on each position qubit is replaced by $R_Y(\pi/2+\epsilon)$, while the coin register and walk operator remain ideal. (a) Success probability $P(\mathbf{x}_0)$ versus the number of steps $t$ for $d=2,3,4,5$ (corresponding to $N=16^2,8^3,8^4,8^5$, respectively). Black dashed lines represent the ideal case, while solid lines correspond to $\epsilon=0.1,0.3,0.5$. (b) Maximum success probability $P_{\max}$ against $N$ for $d=2,3,4,5$.}
    \label{fig:initial_state_error}
\end{figure}

\begin{table}[ht]
\centering
\caption{Comparison of the numerical maximum success probability
$P_{\max}$ with the approximation $\mathcal{F}_0 P_{\max}^{\rm ideal}$
for $d=2,3,4,5$, where
$\mathcal{F}_0=\cos^{2d n_x}(\epsilon/2)$.
The relative reductions are defined as
$\delta_{\rm approx}=1-\mathcal{F}_0$ and
$\delta_{\rm num}=1-P_{\max}/P_{\max}^{\rm ideal}$.
The ideal values correspond to $\epsilon=0$.}
\label{tab:initial_state_fidelity}
\begin{ruledtabular}
\begin{tabular}{cccccccc}
$d$ & $N$ & $\epsilon$ & $\mathcal{F}_0$ &
$\mathcal{F}_0 P_{\max}^{\rm ideal}$ & $P_{\max}$ &
$\delta_{\rm approx}$ & $\delta_{\rm num}$\\
\hline
$2$ & $2^8$ & $0.0$ & $1.000$ & $0.256$ & $0.256$ & $0.000$ & $0.000$\\
    &       & $0.1$ & $0.980$ & $0.251$ & $0.255$ & $0.020$ & $0.004$\\
    &       & $0.3$ & $0.835$ & $0.214$ & $0.229$ & $0.165$ & $0.105$\\
    &       & $0.5$ & $0.603$ & $0.154$ & $0.180$ & $0.397$ & $0.297$\\
\hline
$3$ & $2^9$ & $0.0$ & $1.000$ & $0.374$ & $0.374$ & $0.000$ & $0.000$\\
    &       & $0.1$ & $0.978$ & $0.366$ & $0.373$ & $0.022$ & $0.003$\\
    &       & $0.3$ & $0.816$ & $0.305$ & $0.327$ & $0.184$ & $0.126$\\
    &       & $0.5$ & $0.566$ & $0.212$ & $0.241$ & $0.434$ & $0.356$\\
\hline
$4$ & $2^{12}$ & $0.0$ & $1.000$ & $0.408$ & $0.408$ & $0.000$ & $0.000$\\
    &          & $0.1$ & $0.970$ & $0.396$ & $0.398$ & $0.030$ & $0.025$\\
    &          & $0.3$ & $0.763$ & $0.311$ & $0.318$ & $0.237$ & $0.221$\\
    &          & $0.5$ & $0.469$ & $0.191$ & $0.198$ & $0.531$ & $0.515$\\
\hline
$5$ & $2^{15}$ & $0.0$ & $1.000$ & $0.433$ & $0.433$ & $0.000$ & $0.000$\\
    &          & $0.1$ & $0.963$ & $0.417$ & $0.421$ & $0.037$ & $0.028$\\
    &          & $0.3$ & $0.713$ & $0.309$ & $0.317$ & $0.287$ & $0.268$\\
    &          & $0.5$ & $0.388$ & $0.168$ & $0.175$ & $0.612$ & $0.596$\\
\end{tabular}
\end{ruledtabular}
\end{table}

The quantum spatial search algorithm starts from the uniform initial state $|\Psi(0)\rangle$ defined in Eq.~\eqref{eq:initial_state_valid}. Since imperfect state preparation on quantum hardware may cause deviations from the uniform superposition, we examine how such errors affect the search performance. 

We introduce an over-rotation error $\epsilon$ by replacing each Hadamard gate on the $d n_x=\log_2 N$ position qubits with $R_Y(\pi/2+\epsilon)$, where
\begin{equation}
R_Y\left(\frac{\pi}{2}+\epsilon\right)|0\rangle
=\cos\left(\frac{\pi}{4}+\frac{\epsilon}{2}\right)|0\rangle
+\sin\left(\frac{\pi}{4}+\frac{\epsilon}{2}\right)|1\rangle.
\end{equation}
The same over-rotation $\epsilon$ is applied to all position qubits, while the coin register and quantum-walk operators remain ideal.  

The fidelity between the ideal and perturbed initial states is
\begin{equation}
\begin{aligned}
\mathcal{F}_0
&=|\langle\Psi(0)|\tilde\Psi(0)\rangle|^2\\
&=\cos^{2d n_x}\left(\frac{\epsilon}{2}\right)\\
&=2^{\log_2[\cos^2(\epsilon/2)]\cdot \log_2 N}\\
&=N^{-\kappa(\epsilon)},
\end{aligned}
\label{eq:initial_fidelity}
\end{equation}

where
\begin{equation}
\kappa(\epsilon)
=-\log_2\left[\cos^2\left(\frac{\epsilon}{2}\right)\right]
\simeq\frac{\epsilon^2}{4\ln 2}.
\end{equation}
Thus, $\mathcal{F}_0$ depends only on $N$ and $\epsilon$, independently of $d$ for fixed $N$.

We perform exact state-vector simulations of the quantum spatial search defined by Eq.~\eqref{eq:time_evolv} under periodic boundary conditions.  We consider $\epsilon=0.1,0.3,0.5$.  For Fig.~\ref{fig:initial_state_error}(a), we use $N=2^8,2^9,2^{12},2^{15}$ for $d=2,3,4,5$, respectively.  For Fig.~\ref{fig:initial_state_error}(b), we vary $N$ from $2^6$ to $2^{14}$ for $d=2$, from $2^6$ to $2^{15}$ for $d=3$, from $2^8$ to $2^{16}$ for $d=4$, and from $2^{10}$ to $2^{20}$ for $d=5$. 

Figure~\ref{fig:initial_state_error}(a) shows that the over-rotation error reduces the maximum success probability $P_{\max}=\max P(\mathbf{x}_0,t)$, while the oscillation period remains unchanged. The first peak occurs at approximately the same number of steps, indicating that the optimal number of steps is nearly unaffected by the initial-state error.

To estimate the reduction in the maximum success probability, we decompose the perturbed initial state as
\begin{equation}
|\tilde\Psi(0)\rangle
=\sqrt{\mathcal{F}_0}|\Psi(0)\rangle
+\sqrt{1-\mathcal{F}_0}|\Psi_\perp\rangle,
\end{equation}
where $\langle\Psi(0)|\Psi_\perp\rangle=0$ and $|\Psi_\perp\rangle$ is normalized. After $t$ steps, we obtain
\begin{equation}
|\tilde\Psi(t)\rangle=\sqrt{\mathcal{F}_0}(\hat W\hat R)^t|\Psi(0)\rangle
+\sqrt{1-\mathcal{F}_0}(\hat W\hat R)^t|\Psi_\perp\rangle.
\end{equation}
If the contribution from the orthogonal component and its interference with the ideal component are negligible near the optimal peak, the maximum success probability is approximated by
\begin{equation}
P_{\max}
\simeq \mathcal{F}_0 P_{\max}^{\rm ideal}
=N^{-\kappa(\epsilon)}P_{\max}^{\rm ideal}.
\label{eq:initial_probability_approx}
\end{equation}

We next estimate the tolerable over-rotation error $\epsilon$ required to maintain a given level of the search performance. Under the approximation $P_{\max}\simeq \mathcal{F}_0 P_{\max}^{\rm ideal}$, maintaining $P_{\max}\gtrsim(1-\delta)P_{\max}^{\rm ideal}$, where $\delta$ is the allowable relative reduction in the maximum success probability, requires $\mathcal{F}_0\gtrsim1-\delta$.  

For small $\epsilon$, the fidelity is approximated as
\begin{equation}
\begin{aligned}
\mathcal{F}_0
&=\cos^{2d n_x}\left(\frac{\epsilon}{2}\right)\\
&\simeq\left(1-\frac{\epsilon^2}{8}\right)^{2d n_x}\\
&\simeq\exp\left(-\frac{d n_x\epsilon^2}{4}\right).
\end{aligned}
\end{equation}
Therefore,
\begin{equation}
\begin{aligned}
\exp\left(-\frac{d n_x\epsilon^2}{4}\right)
&\gtrsim 1-\delta,\\
\epsilon^2
&\lesssim-\frac{4\ln(1-\delta)}{d n_x}\\
&\simeq\frac{4\delta}{d n_x}.
\end{aligned}
\end{equation}
Thus, for small $\epsilon$ and $\delta$, the tolerable
over-rotation is estimated as
\begin{equation}
\epsilon
\lesssim 2\sqrt{\frac{\delta}{d n_x}}
=2\sqrt{\frac{\delta}{\log_2 N}}.
\end{equation}
This suggests that the tolerable over-rotation decreases
as $1/\sqrt{\log N}$.  For example, $N=2^{20}$ and
$\delta=0.1$ give an estimated threshold of
$\epsilon\lesssim0.14$.  Table~\ref{tab:initial_state_fidelity} and Fig.~\ref{fig:initial_state_error}(b) compare the numerical maximum success probability with the analytical approximation $P_{\max}\simeq \mathcal{F}_0 P_{\max}^{\rm ideal}$.  The approximation agrees well with the numerical results for $d=4$ and $5$, while larger deviations are observed
for $d=2$ and $3$.

\section{Robustness to Target Location and Spatial Asymmetry}
\label{appdx:robut-spatial-asymmetry}
\begin{figure}[ht]
    \centering
    \includegraphics[width=1.0\linewidth]{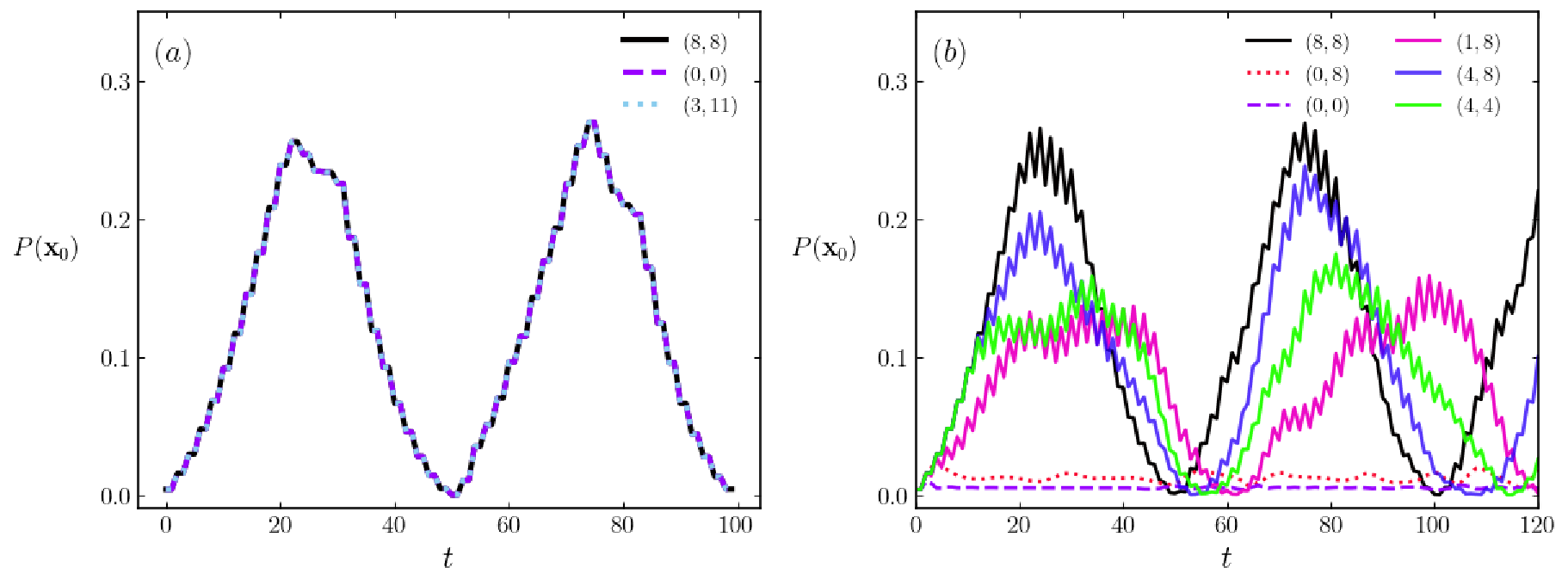}
    \caption{The success probability $P(\mathbf{x}_0)$ at the marked site versus the number of steps $t$ for different marked-site locations $\mathbf{x}_0$ with $d=2$, $N=2^8$. (a) Periodic boundary conditions: the curves for $\mathbf{x}_0=(8,8)$, $(0,0)$ and $(3,11)$ coincide exactly. (b) Non-periodic boundary conditions.}
    \label{fig:target_location_d2}
\end{figure}

We examine how the search dynamics depend on the location of the marked site $\mathbf{x}_0$ for $d=2$ and $N=2^8$, where the vertices are labeled by $\mathbf{x}\in\{0,\dots,15\}^2$.
The initial state is the uniform superposition $|\Psi(0)\rangle$, and all results are obtained by exact state-vector simulation.

Under periodic boundary conditions, the lattice is translationally invariant, so the search dynamics are independent of the marked-site location $\mathbf{x}_0$. This follows because shifting the marked site is equivalent to translating the entire lattice, which leaves the coin and shift operations unchanged.  Fig~\ref{fig:target_location_d2}(a) confirms this property numerically, showing identical success probabilities for three different marked-site locations.

Under non-periodic boundary conditions, shifts beyond the lattice boundary are treated as self-loops, leaving both the site and coin state unchanged. This breaks the translational symmetry. As shown in Fig.~\ref{fig:target_location_d2}(b), the success probability at the center $(8,8)$ is nearly identical to that under periodic boundary conditions.  However, as the marked site approaches the boundary, both the maximum success probability and the oscillation period change. At the boundary, the search performance is significantly reduced, with $P_{\max}=0.023$ at the edge $(0,8)$ and $0.010$ at the corner $(0,0)$.

This reduction can be explained by the existence of a stationary state, similar to the mechanism reported for exceptional configurations~\cite{Ambainis-multi,Nahimovs2016,PhysRevA.94.032334}. When the marked site is located at the boundary, the self-loops allow a stationary state that has a large overlap with the uniform initial state.

Following the stationary-state conditions~\cite{PhysRevA.94.032334}, we construct an unnormalized stationary state
\begin{equation}
|\Psi_s\rangle=|\Psi(0)\rangle-|\delta\Psi\rangle,
\end{equation}
where
$|\delta\Psi\rangle=\frac{k}{m\sqrt{Nk}}
\sum_{l\in B}|\mathbf{x}_0\rangle\otimes|l\rangle$,
and $B$ denotes the $m$ self-loop coin states at the marked site.
This state satisfies $\hat W\hat R|\Psi_s\rangle=|\Psi_s\rangle$.
Thus, the time evolution becomes
\begin{equation}
\begin{aligned}
|\Psi(t)\rangle
&=(\hat W\hat R)^t|\Psi(0)\rangle\\
&=|\Psi_s\rangle+(\hat W\hat R)^t|\delta\Psi\rangle.
\end{aligned}
\end{equation}

The success probability at the marked site $\mathbf{x}_0$ is given by

\begin{equation}
\begin{aligned}
P(\mathbf{x}_0,t)
&=
\sum_{l=0}^{k-1}
|\langle\mathbf{x}_0,l|\Psi(t)\rangle|^2\\
&=
\sum_{l=0}^{k-1}
\left|
\langle\mathbf{x}_0,l|\Psi_s\rangle
+
\langle\mathbf{x}_0,l|
(\hat W\hat R)^t|\delta\Psi\rangle
\right|^2\\
\end{aligned}
\end{equation}

To evaluate its upper bound, we first obtain
\begin{equation}
\langle\delta\Psi|\delta\Psi\rangle=\frac{k}{mN},
\end{equation}
and
\begin{equation}
\sum_{l=0}^{k-1}
|\langle\mathbf{x}_0,l|\Psi_s\rangle|^2
=\frac{k-m}{mN}.
\end{equation}
Using the triangle inequality and the unitarity of $\hat W\hat R$, we obtain

\begin{equation}
\begin{aligned}
P(\mathbf{x}_0,t)
&\leq
\left(
\sqrt{\frac{k-m}{mN}}
+\sqrt{\frac{k}{mN}}
\right)^2\\
&=\frac{(\sqrt{k-m}+\sqrt{k})^2}{mN}\\
&=\mathcal{O}(1/N).
\end{aligned}
\end{equation}
This explains the suppressed search performance at boundary sites observed in Fig.~\ref{fig:target_location_d2}(b).

\section{Quantum circuit for the shift operator}
\label{appdx:shift}

\begin{figure}[ht]
    \centering
    \includegraphics[width=1.0\linewidth]{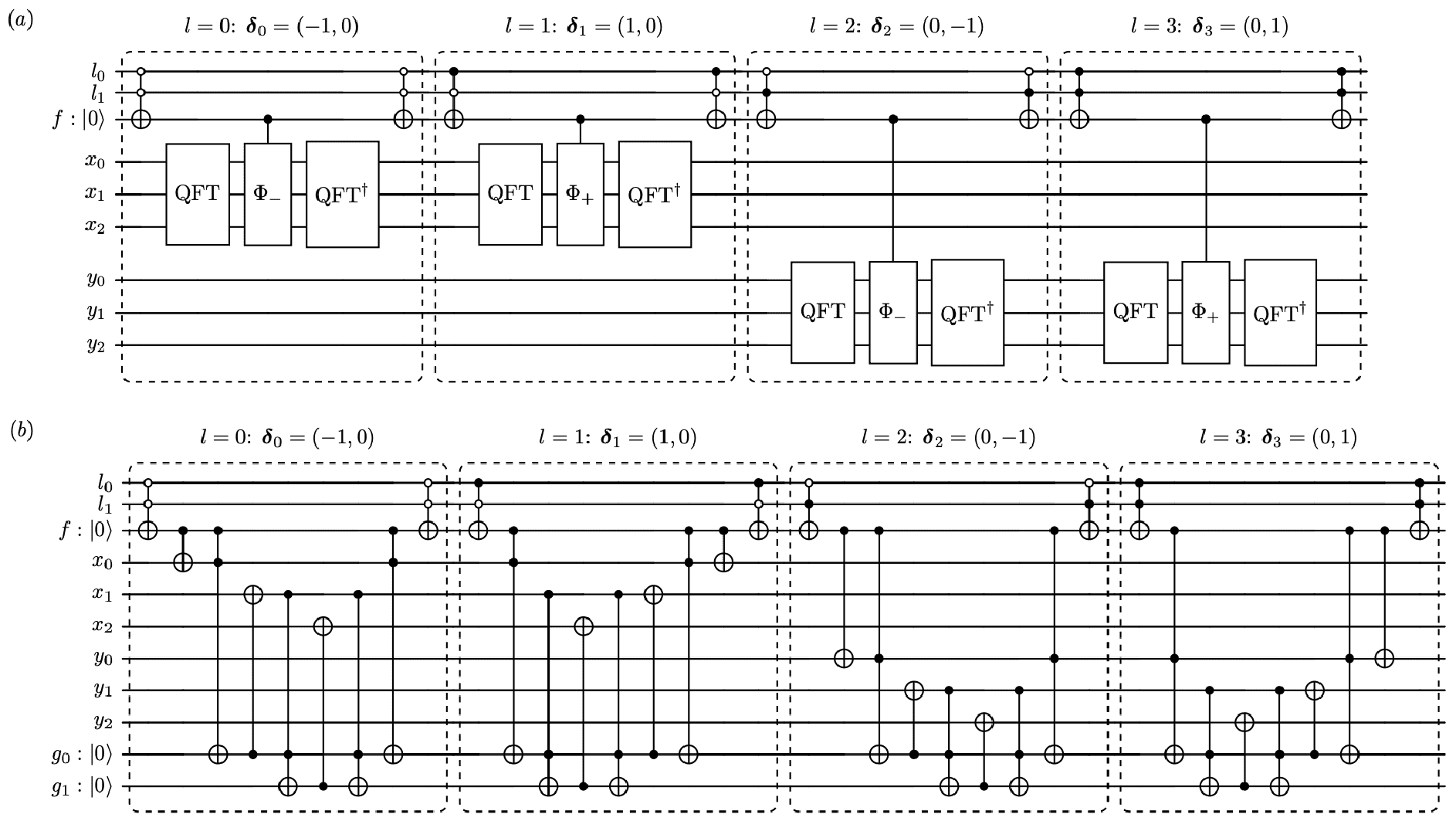}
    \caption{
    Quantum circuits for the shift operator $\hat S_{\rm move}$
    on a two-dimensional square lattice with $n_x=3$.  (a) QFT-based circuit.  (b) Ripple-carry circuit.  Each dashed box implements a translation corresponding to a coin state $\ket{l}$, controlled by the flag qubit $f$.  The qubits $x_i$ and $y_i$ ($i=0,\ldots,n_x-1$) denote the bits of the $x$ and $y$ coordinates, $\Phi_\pm=\bigotimes_{i=0}^{n_x-1}P(\pm\theta_{i})$,
    and $g_0$ and $g_1$ are ancilla qubits.
    }
    \label{fig:shift-qft-ripple}
\end{figure}

\begin{figure}[ht]
    \centering
    \includegraphics[width=0.8\linewidth]{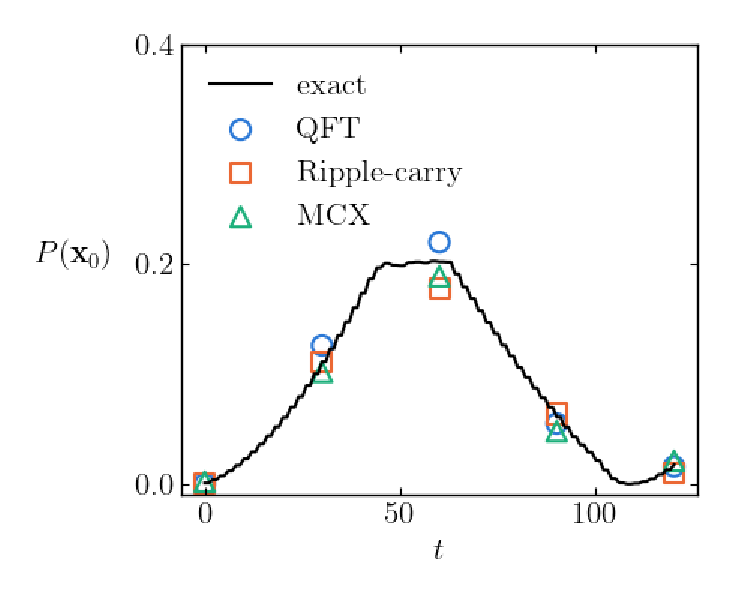}
    \caption{
    Success probability of quantum spatial search on a $d=2$ hypercubic lattice using different shift-operator implementations.  The lattice contains $N=1024$ sites, and the probabilities are calculated from $1000$ measurement shots.
    }
    \label{fig:success-prob-different-circuit}
\end{figure}

The shift operator $\hat S_{\rm move}$ in Eq.~\eqref{eq:shift-op} moves the walker by one lattice site along a coordinate axis determined by the coin state $\ket{l}$.  Each spatial coordinate $x_j$ is encoded in an $n_x$-qubit register.  The shift is implemented by a coin-controlled increment or decrement,
\begin{equation}
    \ket{x_j}\mapsto\ket{x_j\pm1\bmod L}.
\end{equation}
For $L=2^{n_x}$, this modular operation is realized directly as an increment or decrement on the full $n_x$-qubit register. For $L\neq2^{n_x}$, additional operations are required to map $L-1$ to $0$ under the increment and $0$ to $L-1$ under the decrement, so that the walker remains in the valid subspace $\{0,\ldots,L-1\}$. Below, we describe the case $L=2^{n_x}$.

For each coin state $\ket{l}$, a flag qubit $f$ is set to $1$ when the coin register matches $\ket{l}$.  The corresponding increment or decrement is controlled by $f$, which is subsequently uncomputed.  Repeating this procedure for all $2d$ coin states implements $\hat S_{\rm move}$.  Figure~\ref{fig:shift-qft-ripple} shows two implementations based on the quantum Fourier transform (QFT) and
ripple-carry addition.  

\subsection{QFT-based circuit}
\label{appdx:qft-carry}
The QFT-based implementation follows Draper's adder
~\cite{draper2000addition}, which has also been applied
to quantum-walk shift operators~\cite{shakeel2020efficient}.
The QFT transforms the position register into the Fourier
basis, where increments and decrements are implemented
using phase gates.  The operation is expressed as
\begin{equation}
    \mathrm{QFT}^{\dagger}\left(\bigotimes_{i=0}^{n_x-1}P(\pm\theta_{i})\right)\mathrm{QFT},
\end{equation}
where $P(\theta)=\mathrm{diag}(1,e^{i\theta})$ and
$\theta_{i}=2\pi 2^{i}/2^{n_x}$.  The phase gates are controlled by the flag qubit $f$, as shown in Fig.~\ref{fig:shift-qft-ripple}(a).
No additional ancilla qubits are required beyond $f$.  The gate count and circuit depth scale as $\mathcal{O}(n_x^2)$ and $\mathcal{O}(n_x)$, respectively.

\subsection{Ripple-carry circuit}
\label{appdx:ripple-carry}

The ripple-carry implementation is based on Ref.~\cite{cuccaro2004new} and is specialized to
incrementing by $1$.  The circuit propagates carry bits from the least significant bit to the most significant bit using
Toffoli gates and ancilla qubits.  As shown in Fig.~\ref{fig:shift-qft-ripple}(b), the carry bits control the increment operation and are subsequently uncomputed, restoring all ancilla qubits to $\ket{0}$.

The circuit requires $2(n_x-1)$ Toffoli gates, $n_x$ CNOT gates, and $n_x-1$ ancilla qubits.  Both the gate count and circuit depth scale as $\mathcal{O}(n_x)$.  Compared with the QFT-based implementation, the ripple-carry circuit reduces the gate count
from $\mathcal{O}(n_x^2)$ to $\mathcal{O}(n_x)$ at the cost of additional ancilla qubits.

The shift operator consists of $2d$ coin-controlled increments or decrements, each with depth $\mathcal{O}(n_x)$.  For fixed $d$, the total depth therefore scales as $\mathcal{O}(dn_x) =\mathcal{O}(d\log L)=\mathcal{O}(\log N)$.

\bibliography{main}
\end{document}